\documentclass[11pt,a4paper]{article}

\usepackage{jheppub}

\usepackage{rotating}
\usepackage{atlasphysics} 
\usepackage{subfig}
\usepackage{verbatim}     
\newcommand{\bea}{\begin{eqnarray}}
\newcommand{\eea}{\end{eqnarray}}

\usepackage{rotfloat}
\usepackage{multirow}

\usepackage{preprintcover}  
\PreprintCoverPaperTitle{{\boldmath Forward-backward correlations and  charged-particle azimuthal distributions in $pp$ interactions using the ATLAS detector}}  
\PreprintIdNumber{CERN-PH-EP-2012-027}  
\PreprintCoverAbstract{ Using inelastic proton-proton interactions at $\sqrt{s}=$
900~GeV and 7~TeV, recorded by the ATLAS detector at the LHC,
measurements have been made of the correlations between forward and
backward charged-particle multiplicities and, for the first time,
between forward and backward charged-particle summed transverse
momentum.  In addition, jet-like structure in the events is studied by
means of azimuthal distributions of charged particles relative to the
charged particle with highest transverse momentum in a selected
kinematic region of the event.  The results are compared with 
predictions from tunes of the \PYTHIA\ and \HERWIG++ Monte Carlo
generators, which in most cases are found to provide a reasonable
description of the data.}  
\PreprintJournalName{Journal of High Energy Physics}  

\newcommand{\vstrut}{\rule[-7pt]{0pt}{1pt}}   
\newcommand{\pTmin}{$p_{\rm{T min}}$}
\newcommand{\dsumpt}{$\sum {\pt}$}

\newcommand{\rhopt}{\rho_{fb}^{\pt}}
\newcommand{\drhopt}{$\rho_{fb}^{\pt}$}
\newcommand{\ptf}{\sum \pt^f}
\newcommand{\ptb}{\sum \pt^b}

\newcommand{\B}{~\hspace*{-1.7ex}~}
\newcommand{\PYTHIA}{{\sc pythia}}
\newcommand{\PHOJET}{{\sc phojet}}
\newcommand{\HERWIG}{{\sc herwig}}
\newcommand{\GEANT}{{\sc geant}}

\title{\boldmath Forward-backward correlations and  charged-particle azimuthal distributions in $pp$ interactions using the ATLAS detector}

\author{The ATLAS Collaboration}

\abstract{ Using inelastic proton-proton interactions at $\sqrt{s}=$
900~GeV and 7~TeV, recorded by the ATLAS detector at the LHC,
measurements have been made of the correlations between forward and
backward charged-particle multiplicities and, for the first time,
between forward and backward charged-particle summed transverse
momentum.  In addition, jet-like structure in the events is studied by
means of azimuthal distributions of charged particles relative to the
charged particle with highest transverse momentum in a selected
kinematic region of the event.  The results are compared with
predictions from tunes of the \PYTHIA\ and \HERWIG++ Monte Carlo
generators, which in most cases are found to provide a reasonable
description of the data.  }

\begin{document}

\maketitle


\section{Introduction} 
Two general classes of quantum chromodynamic processes combine to
generate the events that are seen in high energy hadron
collisions. The first of these comprises soft processes, which are
commonly pictured in terms of multiple parton interactions at low
transverse momentum, \pT, modelled on multiple $t$-channel gluon
exchanges.  Processes of this kind, together with non-perturbatively
described contributions such as the formation of beam remnants, are
characterised by correlations between numbers of particles per unit of
pseudorapidity over an extended pseudorapidity range~\cite{ref:Kittel}.  At
higher \pT\ values, harder processes arise that are dominated by
perturbatively-described radiation or partonic scatters and are
produced by single- and few-parton exchanges. These can generate
particle jets, which are by definition characterised by short-range
correlation; here high particle multiplicity is found within one unit
in pseudorapidity, quickly diminishing as the pseudorapidity distance
increases. Thus, soft processes produce lower particle multiplicities
with weaker correlations over a wider pseudorapidity distance,
so-called ``long-range correlations'', in comparison to the processes where
a small number of jets are formed. As particle \pT\ values increase
above a few hundred MeV, there is a gradual transition between the
soft regime and the hard. Hence, the correlation measured as a
function of separation in pseudorapidity and of \pT\ probes the
strength of each contribution.

A number of different event generators are available that combine the
basic physical processes appropriately~\cite{ref:Buckley}; however,
the more detailed aspects of the models need to be determined from
data, a process referred to as ``tuning'' the models.  For example,
there are choices to be made as to whether the initial- and
final-state gluon radiation are interleaved with the multiple parton
interactions (MPI) and whether the modelling of the showers is to be
ordered in terms of \pT, angle, or the momentum transfer squared,
$Q^2$, of the parton scatter.  String or cluster fragmentation may be
selected, and diffractive processes should be taken into
consideration. Factors such as these can affect the correlations
between the generated particles.

The tunings of the non-diffractive models have made use of output from many
experiments; a variety of event characteristics were taken into
account, such as distributions in particle multiplicity, transverse
momentum and rapidity. A number of different tunes of the
leading-order plus leading-logarithm generator
\PYTHIA~\cite{ref:Pythia, ref:Pythia6, ref:Pythia8} exist
that provide a reasonably good description of events up to
LHC energies~\cite{ref:Pythia-tuning,ref:professor,ref:atlastune}.
The \PYTHIA\ model uses string
fragmentation~\cite{ref:strings1,ref:strings2}, in contrast to the
cluster fragmentation used in the \HERWIG\
model~\cite{ref:herwigcomp,ref:Herwig21,ref:Herwig25}.

The correlations in pseudorapidity between the particles in an event
have not been explicitly considered in the existing Monte Carlo (MC)
tunes, and provide a further means of improving and helping to
discriminate between the models~\cite{ref:KWPS}.  Such correlations
are the subject of the first part of the present study, which was
carried out using proton-proton collisions recorded with the ATLAS
detector~\cite{ref:AtlasExp} at the LHC.\footnote{ATLAS uses a
right-handed coordinate system with its origin at the nominal
interaction point (IP) in the centre of the detector and the $z$-axis
along the beam pipe. The $x$-axis points from the IP to the centre of
the LHC ring, and the $y$-axis points upward. Cylindrical coordinates
$(r,\phi)$ are used in the transverse plane, $\phi$ being the
azimuthal angle around the beam pipe. The pseudorapidity is defined in
terms of the polar angle $\theta$, with respect to the $z$-axis, as
$\eta=-\ln\tan(\theta/2)$.  Throughout the paper the term ``forward''
is used to refer to the direction of positive $z$ or $\eta$, and
``backward'' to refer to the negative direction. } The measurements
were inspired by earlier work by the UA5 experiment~\cite{ref:UA5} and
follow published studies at the LHC of the multiplicities
themselves~\cite{ref:MB20,ref:CMSmult,ref:LHCbmult,ref:ALICEmult1,ref:ALICEmult2}.
They also complement an earlier analysis by CMS of charged-particle
correlations~\cite{ref:CMScorr} and a recent analysis by ATLAS of
two-particle correlations~\cite{ref:ATL2PC}. The approach taken here
is to choose pairs of pseudorapidity intervals equal in size and
symmetrically located in the forward and backward direction, allowing
a measurement of what will be termed a forward-backward (FB)
correlation between quantities in the two intervals.  To determine the
effects of both short-range and long-range physical correlations, the
FB correlations are measured for increasing separation between the two
selected pseudorapidity intervals.

Two types of FB correlation are presented: correlations in
charged-particle multiplicity and correlations in the summed values of
the transverse momenta of the charged particles.  A further
perspective comes from increasing the value of the minimum transverse
momentum, \pTmin, of the particles that are considered. In this way
the effects of the soft processes can be gradually reduced, while the
effects of the harder processes remain.  The use of data taken at two
very different proton-proton centre-of-mass energies, $\sqrt{s}=
900$~GeV and 7~TeV, gives further information that can be compared
with MC tunes based on $p\bar{p}$ interactions at the intermediate energy of
the Tevatron, at $\sqrt{s}= 1.96$~TeV.  The 7~TeV data provide access
to a new energy regime, which is the main goal of the LHC. It is
important to understand the relationship between softer and harder
processes in these two regimes in order to confirm our understanding
of the background processes at the energies where new physics might be
found.

Although such correlation measurements are
sensitive to the presence of jets in events, no jet features are explicitly
sought or exploited. Jets are by definition
associated with short-range correlations in both pseudorapidity and
azimuthal angle, and to obtain additional information about these correlations,
a further analysis was performed using a different
technique. Since the direction of a jet is often close to that of the
highest-\pT\ particle in the event, the charged particle with the highest
\pT\ is identified and the distributions of the other charged
particles are studied as a function of their azimuthal angle relative to its
direction.

The distribution of particles in the azimuthal plane of an event has
been studied before at Tevatron energies~\cite{ref:dphi6, ref:dphi7,
ref:dphi8} and at the LHC~\cite{ref:dphi9}.  Taken relative to the
highest-\pT\ particle, the azimuthal distribution of particles shows
two peaks, centred at zero and at $\pi$, that can indicate jet structure.
In the events studied in the present measurement, however, these peaks
are typically not observed in single events, most of which contain too
few hard particles to form clearly defined jets, but become visible
statistically when assembling the distribution over many events. The
particle distributions also display the so-called ``underlying
event'', the effects of which are particularly apparent at azimuthal angles
perpendicular to the jets~\cite{ref:dphi9}. However, the principal
goal of the present study is an understanding of the shape of the
first jet-like peak, which is described differently depending on the
MC  model used. Again, it is found that the transition from
900~GeV to 7~TeV LHC energies provides a significantly different
perspective on the physics.

This paper is structured as follows. The statistical methods for the
correlation analysis and the azimuthal analysis are described in
sections \ref{sec:statmeth} and \ref{sec:dphi}.  Section \ref{sec:app}
describes the ATLAS detector and the event selection, while the MC
samples are described in section \ref{sec:MonteCarlo}. Details of the
analysis method are given in section \ref{sec:analcorr}. The
systematic uncertainties are discussed in section
\ref{sec:systematics}, and the results and conclusions are presented
in sections \ref{sec:results} and \ref{sec:conclusions}.

\section{Forward-backward correlations}
\label{sec:statmeth}

All the correlations measured in this study, whether from symmetric or
asymmetric pseudorapidity intervals, are referred to as
forward-backward (FB) correlations and are measured using charged
particles only.  The FB multiplicity correlation, $\rho^n_{fb}$,
between two particle multiplicities is the normalised
covariance between the two distributions, relative to the mean value
of each. For a given data sample or MC sample, it is
defined in the usual way as:
\begin{equation}
\rho^n_{fb}=\frac{\langle (n_f -\langle n_f \rangle )(n_b- \langle n_b\rangle )\rangle }
{\sqrt{\langle (n_f-\langle n_f\rangle )^2\rangle \langle (n_b-\langle n_b\rangle )^2\rangle }}\,
=\, \frac{\sum x^n_f x^n_b}{N \sigma^n_f \sigma^n_b}.
\label{eqn:bcorr}
\end{equation}
Here, $n_f$ and $n_b$ are the respective multiplicities of particles
of interest in two chosen forward and backward intervals in an event,
$\langle\,\,\rangle$ denotes a mean over the events in the sample, $N$
is the total number of events, $x^n_f,\, x^n_b$ are the differences
between $n_f$,\, $n_b$ and their means, and $\sigma^n_f,\,\sigma^n_b$
are the standard deviations of $n_f$ and $n_b$ about their means. The
sum in the right-hand side of the equation is taken over the events in
the sample.  In the present measurement, particles of interest are
those above a given \pTmin\ value.  Intervals in pseudorapidity of
size $\delta\eta = 0.5$ were chosen, which allows good statistical
accuracy on each measured point while preserving a sensitivity to
physically interesting variations with pseudorapidity $\eta$. A range
$-2.5 < \eta < 2.5$ was considered, corresponding to the inner
tracking detector acceptance, divided into five forward and five
backward intervals.

The FB momentum correlation, \drhopt, between two summed transverse
momentum (\dsumpt) values is defined similarly:
\begin{equation}
\rhopt=\frac{\langle (\ptf -\langle \ptf \rangle )(\ptb- \langle \ptb
  \rangle )\rangle }{\sqrt{\langle (\ptf -\langle \ptf \rangle
    )^2\rangle \langle (\ptb -\langle \ptb \rangle )^2\rangle }}
= \frac{\sum x^{\pt}_f x^{\pt}_b}{N \sigma^{\pt}_f \sigma^{\pt}_b},
\label{eqn:bcorrp}
\end{equation}
where $\ptf$ and $\ptb$ are the sums of the  absolute transverse momentum
values of the charged particles in the two chosen forward and backward
intervals in an event, $x^{\pt}_f,\, x^{\pt}_b$ are the differences
between $\ptf$,\, $\ptb$ and their means, and
$\sigma^{\pt}_f,\,\sigma^{\pt}_b$ are the respective standard
deviations. The \pTmin\ values used in measuring the correlations are
discussed in the sections below.

\section{Azimuthal distributions}
\label{sec:dphi}

To obtain sensitivity to the presence of jets and jet-like structures
in the events, the highest-\pT\ charged particle in a given $|\eta|$
region in an event is identified and termed the ``leading
particle''.  The azimuthal difference $\Delta\phi$ between the leading
particle and any other accepted charged particle is the unsigned angle
(in the range 0 -- $\pi$) between the two particles in the transverse
plane. The variable $N^T$ is defined as the number of selected charged
particles, taken over the entire event sample, in a given interval of
$\Delta\phi$ of width $\delta\phi$, where $\delta\phi$ takes
the value $\pi/50$ in the present analysis. A \pT\ $>$ \pTmin\
requirement is again imposed on the selected particles.

\begin{figure}[!ht]
\centering
  \includegraphics[width=0.48\textwidth]{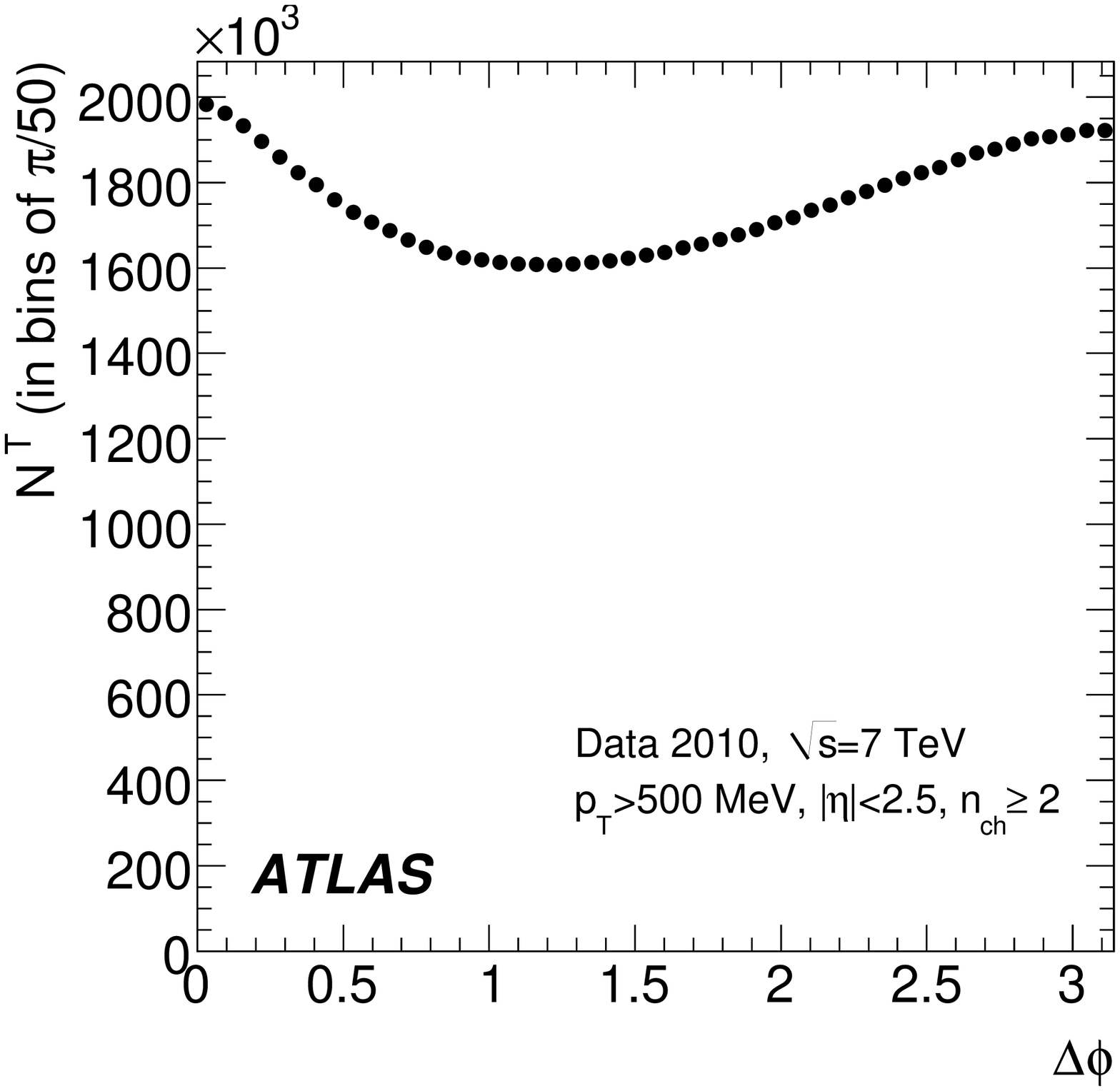}
  \includegraphics[width=0.48\textwidth]{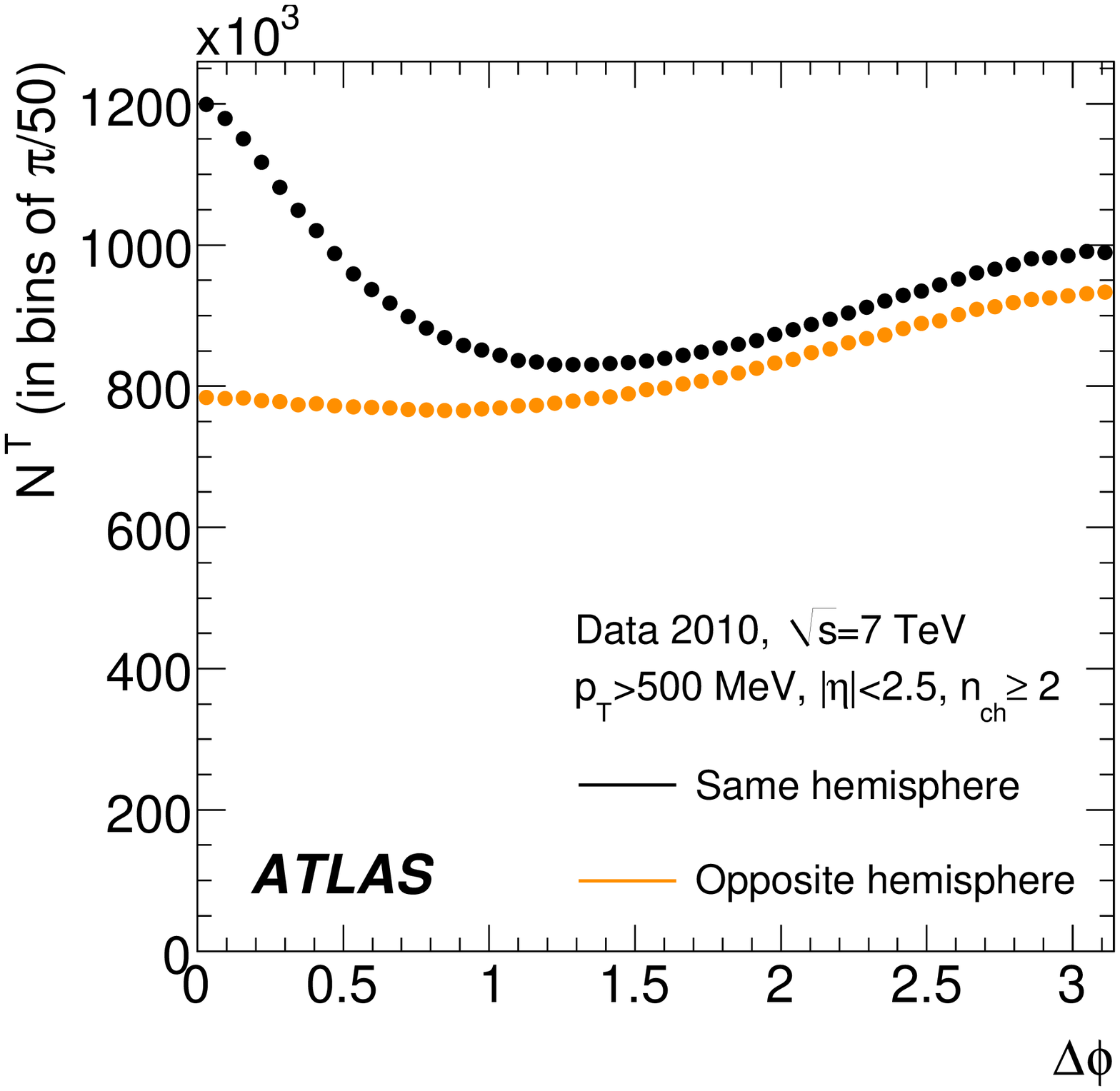}
\\[-1.9ex]\hspace*{3.5ex}(a)\hspace*{0.45\textwidth}(b)\\[1.5ex]
\caption{Azimuthal distributions for tracks with \pTmin\ = 500~MeV and
$|\eta| < 2.5$ in data at $\sqrt{s}=7$~TeV.  (a) Uncorrected
distribution of $\Delta\phi$; (b) distributions for the same
hemisphere in $\eta$ as the leading track, and for the opposite hemisphere.
The statistical uncertainties are smaller than the plotted symbol.}
\label{fig:CCraw}
\end{figure}

In the general case of jet production, it is expected that the
distribution in $N^T$ should show peaks at $\Delta\phi = 0$ and at
$\Delta\phi = \pi$, with a minimum lying in between.  As observed in
the present study (figure~\ref{fig:CCraw}a), this minimum takes the
form of a flat contribution, or pedestal, which arises from the
overall charged-particle activity in the event~\cite{ref:dphi9}
together with uncorrelated backgrounds that are hard to model, such as
fake tracks in the detector. The present study focuses on the peak
structure, suppressing the influence of the pedestal on the
measurement by performing a subtraction. Since the pedestal
subtraction is performed in a corresponding way on the MC models
tested, the latter do not need to model the pedestal level accurately
for this analysis. In each case, we fit the observed $N^T$
distribution with a quadratic polynomial, determine the minimum value
$N^T_{min}$, and subtract this value from all the $N^T$ values in the
distribution. The ``subtracted'' number of charged particles per
radian in azimuthal angle is then normalised to give the quantity
\begin{equation}
N^T_{sub} = \frac{(N^T - N^T_{min})}{\sum(N^T - N^T_{min})}\,,  
\end{equation}
where the sum is taken over the bins of $\delta\phi$ used in the
analysis.

Averaged over many events, the $N^T_{sub}$ variable is sensitive to
the presence of jet pairs whenever both jets occur within the accepted
$\eta$ interval. The direct effect of the second jet is removed,
suppressing pedestal effects in an alternative way, by examining a
second variable.  We distinguish between particles produced in the
same forward-backward event hemisphere in $\eta$ as the leading
particle, and those produced in the opposite hemisphere
($\eta*\eta_{lead}\ge 0$ and $<0$, respectively), denoting these as
$N^T_{same}$ and $N^T_{opp}$ respectively, as shown in
figure~\ref{fig:CCraw}b.  The normalised asymmetry is then defined as
\begin{equation}
 N^T_{SO} = \frac{(N^T_{same} - N^T_{opp})}{\sum(N^T_{same} - N^T_{opp})}\,.
\end{equation}
On the assumption that the leading particle indicates the possible
presence of a jet, $N^T_{SO}$ is expected to take large positive
values for $\Delta\phi\approx 0$ but to be smaller for larger
$\Delta\phi$. Thus the jet structure associated with the leading particle is
well-characterised, but the effects of the opposing jets tend on average to
cancel out. Any non-zero value at large $\Delta\phi$ reflects
long-range correlations in $\eta$ with jet effects largely
removed.  In this way, the two variables enhance the sensitivity to
different aspects of jet-like events, and are sensitive to an aspect
of long-range correlations that complements the FB measurements.

\section{Detector and event selection}
\label{sec:app}

The ATLAS detector at the LHC covers almost the
entire solid angle around the collision point with layers of tracking
detectors, calorimeters and muon chambers.  In the present analysis,
tracks and vertices are reconstructed using the inner detector, which
consists of a silicon pixel detector, a silicon strip detector and a
transition radiation tracker. These extend in total from radii of
50.5~mm to 1066~mm from the beam line, and are located in a 2~tesla
axial magnetic field provided by a superconducting solenoid. A
track from a charged particle has typically 11 silicon detector
hits and more than 30 transition radiation tracker hits.  Coverage is
obtained in the pseudorapidity range $|\eta| < 2.5$.

  The data analysed here were collected with the ATLAS detector in
2009 at $\sqrt{s}=$900~GeV, and in early 2010 at $\sqrt{s}=$7~TeV,
corresponding to integrated luminosities of 7~$\mu$b$^{-1}$ and
190~$\mu$b$^{-1}$ respectively. The maximum instantaneous luminosity
was $\sim1.9\times 10^{27}$~cm$^{-2}$s$^{-1}$ (at 7~TeV), and the
probability of additional interactions in the same bunch crossing as a
given event was of the order of $10^{-3}$.

This analysis is based on the same dataset as was used for the ATLAS
minimum-bias particle multiplicity analysis~\cite{ref:MB20} and uses
the same minimum-bias scintillator trigger, the same event selections,
and the same quality criteria on the track reconstruction and
selection.  An event in this dataset was required to have at least one
primary vertex, formed from a minimum of two tracks with \pT\ $>$
100~MeV and consistent with the beamline.  The primary vertex with
tracks giving the highest $\Sigma p_T^2$ was selected, and events with
a second primary vertex with four or more tracks were rejected, to
minimise any possible effects due to pile-up. Tracks with an impact parameter
above 1.5 mm in the transverse plane were removed to reduce the
effects of particles from secondary decays.  Around ten million events
at 7~TeV and one million events at 900~GeV were selected at this
stage.

For the FB correlation analysis, event samples were formed
having at least two tracks in the event selected with \pT\ $>$ \pTmin\
and $|\eta|<$ 2.5, for a range of \pTmin\ values: 100, 200, 500~MeV,
1.0, 1.5, 2.0~GeV (see section \ref{sec:fbresults}).
For the azimuthal analysis, a value of \pTmin = 500~MeV was used in the
final track and event selection, as discussed in section
\ref{sec:analaz}.


\section{Monte Carlo models}
\label{sec:MonteCarlo}

It was necessary to apply corrections to the observed FB correlations
in order to obtain the final results. These corrections were evaluated
using simulated events generated with the MC09 tune of
\PYTHIA~6~\cite{ref:Pythia6, ref:MC09}. This tune, based on
\PYTHIA~6.4.21, was created by ATLAS to emulate the particle
production at the Tevatron and provide predictions for the LHC. In
\PYTHIA~6, the initial-state gluon radiation is interleaved with the
MPI, which is expected to reduce the difference in strength between
short- and long-range correlations. Four further tunes, chosen to make
use of different physical approaches as described below, were employed
to evaluate model-dependent systematic effects.
Around 800,000 MC events for each tune were used.  Each of these
samples contained a mixture of non-diffractive, single-diffractive and
double-diffractive events.  The subprocesses were combined according
to the generator cross-sections.
All the MC samples were passed through standard ATLAS software 
programs~\cite{ref:atsim}, based on \GEANT4~\cite{ref:geant}, to
simulate the trigger and detector effects before event reconstruction.

 Two of the further tunes were also based on \PYTHIA~6.4.21.  The 
DW tune~\cite{ref:DW} was created to describe underlying-event (UE) data at the
Tevatron and uses a $Q^2$-ordered shower, in contrast to the \pT\
shower ordering used by the other \PYTHIA\ tunes. In the 
Perugia0~\cite{ref:Perugia0} tune the soft contribution was optimised using
minimum-bias data from the Tevatron and the CERN Sp\=pS
colliders. A \PYTHIA~8 sample~\cite{ref:Pythia8} was also used, with
default parameter values (Tune 1). In \PYTHIA~8, the final-state
radiation (FSR) as well as the initial-state radiation (ISR) are
interleaved with the MPI~\cite{ref:interleaved}. An alternative approach was
provided by using \PHOJET~1.12.3.15~\cite{ref:Phojet}. This model
uses a dual-parton event description, with soft hadronic processes
described by pomeron exchange and semi-hard processes described by
perturbative parton scattering. It has similar fragmentation and
hadronisation to the \PYTHIA\ model, and includes an additional central
diffractive component.

 The final results are compared with a selection of the above tunes and
additional tunes, some of which have made use of LHC minimum-bias
and underlying-event data.  None of the tunes has taken account of FB
correlation measurements. In general, the effect of the newer tunes
has been to change the numbers of multiple parton interactions and the
degree of colour reconnection.  Perugia2011~\cite{ref:Perugia0} is a
more recent tune of \PYTHIA~6.4 that uses the usual \pT-ordered shower
model, the interleaved MPI model, and a new set of colour
reconnections.  AMBT2B is a tune of \PYTHIA~6.4 to ATLAS minimum-bias
data~\cite{ref:AMBT2B}. The new \PYTHIA~8 4C tune was 
used~\cite{ref:Pythia8}, and also
\HERWIG++~\cite{ref:Herwig21,ref:Herwig25}, in versions 2.1 and 2.5,
tuned to minimum-bias LHC data at 900~GeV (tune MU900-2) and to
the underlying event at 7~TeV (tune UE7-2). The \HERWIG\ model uses
an angular-ordered parton shower and cluster fragmentation, but does
not include diffractive processes, which constitute approximately 35\%
of the \PYTHIA\ cross section and 20\% of that of \PHOJET.
Appendix~\ref{app:a} presents in tabular form the main features of the
\PYTHIA\ and \HERWIG\ tunes that have been used.

\section{Analysis method}
\label{sec:analcorr}
The correlations and distributions of physical interest are defined at
the so-called ``particle level'' of the events; that is to say, in
terms of the stable charged particles produced in the primary
interaction with lifetime $c\tau >10$~mm.  The kinematic region
in a given measurement was defined by a requirement that there should
be at least two such particles associated with the event,
having \pT\ $>$ \pTmin\ and $|\eta|$ within a specified range up to a
maximum value of 2.5.  The FB correlations and the azimuthal
distributions observed in the detector were corrected to the
applicable kinematic region at the particle level.

\subsection{Forward-backward correlations}
\label{sec:analfb}

As observed in the detector, the FB correlations are modified by
detector effects such as track reconstruction efficiency, trigger efficiency,
vertex finding efficiency, and the detection of secondary particles
arising from decays and interactions in the detector material.  A
simple weighting-factor correction to the number of measured tracks in
a given interval (or to their summed \pt) will not suffice when
correlations are being measured, because the corrections may
themselves be correlated between the different intervals.  In the
context of a set of variables with multiple correlations
between them, a natural method to use for determining the correlations
is multiple regression theory~\cite{ref:stats}.  The following multiple regression
technique was used to apply corrections to the observed FB
correlations so as to obtain the FB correlations at the particle
level.

In general, given four statistically varying quantities that may be
correlated among themselves, the linear regression relationships take the
form 
\bea 
x_1=b_{12.34}~x_2+b_{13.24}~x_3+b_{14.23}~x_4\,\nonumber  \label{equ:sys_lin_reg}\\
x_2=b_{21.34}~x_1+b_{23.14}~x_3+b_{24.13}~x_4\, \nonumber\\
x_3=b_{31.24}~x_1+b_{32.14}~x_2+b_{34.12}~x_4\, \nonumber\\
x_4=b_{41.23}~x_1+b_{42.13}~x_2+b_{43.12}~x_3. 
\eea 
Each variable $x_i$ denotes the difference between a statistically
varying quantity and its sample mean. In the present case, this
notation refers to track or charged-particle multiplicity, or to a
summed transverse momentum value.  In a given sample of events, each
partial regression coefficient $b_{ij.kl}$ denotes how a chosen
statistical quantity $x_i$ varies with one of the others, $x_j$, on
average, for fixed values of the other two. A system of normal
equations can be solved to determine these coefficients from the event
sample.

The indices 1 and 2 are here used respectively to denote forward and
backward particle-level quantities, and the indices 3 and 4
to denote forward and backward observed quantities,
as measured in the detector. Each quantity
may be correlated with any of the others.  Particularly strong
partial correlations exist between $x_1$ and $x_3$, of course, and
likewise between $x_2$ and $x_4$.  The
regression coefficients are determined by the use of MC
event samples in which both the particle-level and the simulated 
detector-level quantities are known.

In order to link a FB correlation at the particle level
$\rho_{part}\equiv\rho_{12}$ to the corresponding observed FB
correlation $\rho_{obs}\equiv\rho_{34}$, we start from the second and
third lines of (\ref{equ:sys_lin_reg}).
These equations are multiplied by $x_3$
and by $x_2$, respectively, and after using the definitions of
equations (\ref{eqn:bcorr}) and (\ref{eqn:bcorrp}) we obtain:
\bea
\rho_{part}= \alpha + \beta \cdot \rho_{obs} \label{equ:b_obs_b_tru},
\eea
where
\bea
\alpha=\frac{b_{23.14}\sigma_3^2 - b_{32.14}\sigma_2^2 
+ b_{21.34}\sigma_1\sigma_3\rho_{13} -b_{34.12}\sigma_2\sigma_4\rho_{24}}{b_{31.24}\sigma_1\sigma_2}, 
~~~~~~~~~~\beta=\frac{b_{24.13}\sigma_3\sigma_4}{b_{31.24}\sigma_1\sigma_2}. 
\label{eqn:alpha_beta}
\eea 
The particle-level correlations are therefore linearly
related to the observed correlations and may be obtained from them by
means of the two coefficients $\alpha$ and $\beta$.

The relationships between $x_1$ and $x_3$, and between $x_2$ and
$x_4$, are due largely to the effect of track reconstruction efficiency;
these relationships are contained in the coefficient $\beta$. The
other terms contain detector effects such as trigger and vertex
finding efficiencies, and also effects of physical FB correlations, as
modelled in the MC simulations.  These affect the coefficient
$\alpha$, where they partially cancel. The values of $\alpha$ and
$\beta$ were calculated using a range of MC tunes together with the
detector simulation, and the variations on the values obtained then
give rise to a systematic uncertainty on the results. This uncertainty has been
reduced to a low level by this method and is contained mainly within
the coefficient $\alpha$, whose value was found to lie in the range
$0.07\pm0.03$. The value of $\beta$ lies in the range 0.96 -- 0.97
(0.97 -- 0.98) for the multiplicity (summed \pT) measurements. 
The full set of systematic uncertainties
on the results is discussed in section~\ref{sec:systematics}.


To validate the above approach, the linearity of the relationship
\ref{equ:b_obs_b_tru} was checked for both MC and data. It was verified
from the MC samples that the analysis procedure regenerated the
correlations that were calculated from the generated particles
directly.  The data sample did not include events produced with fewer
than two reconstructed tracks associated with the primary vertex; a
correction of up to 0.9\% was applied for the small bias associated
with this.

\subsection{Azimuthal distributions}
\label{sec:analaz}

The azimuthal analysis employed the same track and event selection as
the FB correlations analysis (section \ref{sec:app}) but imposed a
higher minimum transverse momentum requirement of \pTmin=500~MeV on
accepted tracks. This minimises systematic uncertainties by selecting
a kinematic region with track reconstruction efficiencies that are
flat in \pT, a procedure made necessary by the subtraction methods
used in this section of the analysis.  Histograms of $N^T_{sub}$ and
$N^T_{SO}$ were made as a function of $\Delta\phi$, and required
corrections for detector effects in order to obtain the particle-level
distributions.

One set of effects comes from a loss of non-leading tracks due to
detector inefficiency, and contamination from non-primary tracks. To
correct for these effects, each non-leading track was weighted by a
factor $(1 - b)/\epsilon$, where $b$ represents the fraction of
background due to non-primary tracks and $\epsilon$ is the mean track
reconstruction efficiency for the selected tracks, evaluated as a
function of \pT\ and $\eta$.  These terms were evaluated from 
MC simulation.  The normalisation procedure gives a large
cancellation of the effects of non-leading track reconstruction
inefficiency, while preserving the shape of the variations in
$\epsilon$; these are more important than the average value.

The loss of a leading track, due to track reconstruction inefficiency,
may lead to the loss of the event or else to the wrong identification
of a non-leading track as the leading track. A correction was applied
to the observed distributions of figure \ref{fig:CCraw} to remove the
shape distortions caused by these effects. It was done in three steps.
First, in a given \pT\ and $\eta$ range, the fraction
$\epsilon_l$ of events that have the correct leading track was estimated by
integrating the track reconstruction efficiency (determined from
MC-simulated events) over the observed \pT\ and $\eta$ spectrum of the
events in the data.  Second, starting from the observed data
distribution in $\Delta\phi$, a series of new distributions was constructed 
by removing an increasingly large random fraction $f$ of the
observed leading tracks from the data sample.  The
resulting shape distortion was found to be linear in $f$ in each
$\Delta\phi$ bin. This made it possible to extrapolate the original
data distribution to what it would have been if $\epsilon_l$ were
100\%.  The correction to the final results is up to 2\%.  

The distributions of $N^T_{sub}$ and $N^T_{SO}$ were made for three
ranges of $\eta$, namely $|\eta| <$ 1.0, 2.0 and 2.5, in order to provide 
a broad-based comparison with the MC models.

\section{Systematic uncertainties}
\label{sec:systematics}

\subsection{Forward-backward correlations}

The largest systematic uncertainty on the results 
arises from the uncertainty on the track reconstruction efficiency. A smaller
effect arises from the variation of the correction parameters $\alpha$
and $\beta$ with the MC model used to obtain them (c.f.~section
\ref{sec:analfb}).  The effects from the uncertainties on trigger and
vertex finding efficiencies are small.

In this section, uncertainties are given as percentages of the central values
and symmetric positive and negative uncertainties are taken.
The model-dependence of the analysis was evaluated for the 7~TeV data,
and the conclusions were also applied to the 900~GeV data. The
chosen initial set of five MC tune samples were used to correct the
data, and the spread in the results relative to the standard
MC09, of the order of $ 1\%$, was taken as the systematic uncertainty due
to the model-dependence of the correction procedure.  In addition, the
diffractive components of the MC09 sample were varied by $\pm10\%$;
this changed the measured correlations by typically $\pm 0.8\%$ of their value.

The systematic effects on the FB correlations due to uncertainties on
the trigger, vertex finding and track reconstruction efficiencies were
studied by varying these efficiencies in the MC, and then
re-evaluating the FB correlations using the standard correction
parameters.  Uncertainties on efficiencies of $0.9$\% ($1.6$\%) and $0.4$\%
($0.6$\%) were used at 7~TeV (900~GeV) for the trigger and vertex
finding, respectively, and a track reconstruction uncertainty varying
between $2$\% and $15$\%~\cite{ref:MB20} was used. The systematic
variations on the standard selection were achieved by removing a
fraction of the events or tracks in a sample, randomly selected, according to the
uncertainties on the respective efficiencies. This simulated a
decrease in the efficiency which was taken as the
corresponding systematic uncertainty.

In the case of the trigger and vertex efficiencies, the deviations on
the FB correlations are  $\leq0.2\%$ and $\sim0.2\%$, respectively. When
the number of tracks in the events was varied as described above, the
resulting FB correlation varied by $1.2\%\,(1.4\%)$ in the central
pair of $\eta$-intervals, increasing to $3.3\%\,(3.0\%)$ in the
outermost pair for multiplicity (momentum). 

The total systematic uncertainty was obtained by summing the
individual contributions in quadrature, and was combined in quadrature
with the statistical uncertainty to obtain the total uncertainty.  The
total systematic uncertainty is dominated by the errors on the model
dependence and on the track reconstruction efficiency, and both of
these are highly correlated between the two energies.  In taking the
ratios of the correlations at different energies, the systematic
uncertainties are therefore small relative to the statistical and can
be neglected.

\subsection{Azimuthal distributions}

Performing the subtractions and the normalisations reduces the
systematic uncertainties on the $\Delta\phi$ distributions
considerably. Of primary importance is the uncertainty on the shape of the
distribution, quantified here by estimating the relative
uncertainty on the first bin in $\Delta\phi$.  The most important
effect comes from the event selection efficiency (1\% in the 900~GeV
data and 3\% in the 7~TeV data), and mainly affects events with a
small number of tracks. Other important effects come from the
correction procedure, which contributes 2\%, and the effects of the
track momentum resolution in the vicinity of the boundaries of the
kinematic acceptance.  The latter was estimated by varying the values
of the cuts on \pT\ and $\eta$ by their resolution and repeating the
analysis.  This gave variations of 1--2\% and 0.2--0.5\% from \pT\ and
$\eta$ respectively.

A variety of additional effects gave uncertainties at the 0.1--0.3\%
level.  These comprised errors in the choice of leading charged
particle (estimated from MC), uncertainty on the fraction of leading
particles that were not reconstructed (also estimated from MC), and
uncertainties on the track reconstruction efficiencies of the
non-leading tracks. The uncertainty on the track reconstruction
efficiency, measured to be on average 15\%~\cite{ref:MB20}, was
propagated to the non-leading track correction. It corresponds to a
relative systematic uncertainty on the distribution of 0.1--0.2\%.  A
final contribution of similar magnitude came from uncertainties on the
number of secondary particles, depending on the choice of impact
parameter cut.  The track reconstruction efficiency difference between
the forward and backward halves of the detector was found to have a
negligible effect.

The total systematic uncertainty was obtained by summing the individual
contributions in quadrature, and was combined in quadrature with the
statistical uncertainty to obtain the total
uncertainty.  The statistical uncertainties dominate the
measurements at 900~GeV, but in the 7~TeV data the statistical and
systematic uncertainties are comparable.  


\section{Results}
\label{sec:results}

\subsection{Forward-backward correlations}
\label{sec:fbresults}

The observed FB correlations were corrected to the particle level,
using the procedure described in section \ref{sec:analcorr}, and all
results are quoted at this level.  The correction typically increased
the correlation value above the observed, uncorrected value by
0.05--0.08, in both multiplicity and momentum.

\begin{figure}[!ht]
\centering
  \includegraphics[width=0.75\textwidth]{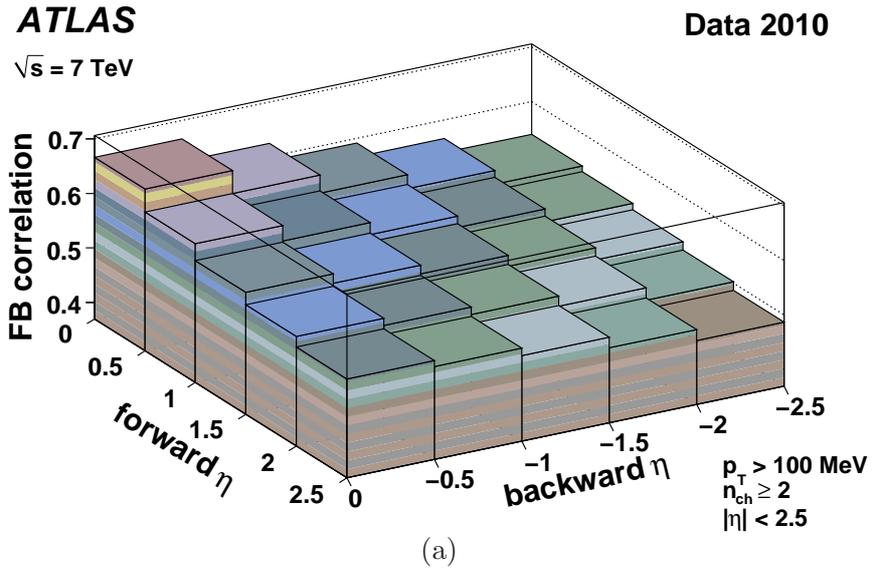}
\\[0.05\textwidth]
 \includegraphics[width=0.75\textwidth]{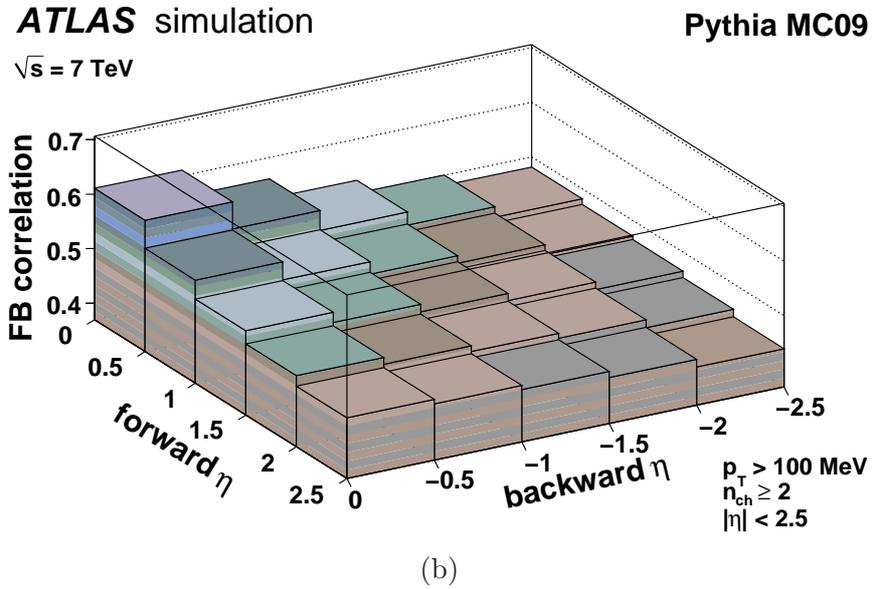}
\\[-0.54\textwidth](a)\\[0.51\textwidth](b)\\[0.0\textwidth]
\caption{Multiplicity correlation at 7~TeV in a matrix of forward/backward
$\eta$-intervals for $|\eta| < 2.5,$ for events with at least
two charged particles with $p_{\rm T} > 100$~MeV and
$|\eta| < 2.5$. (a)~data; (b)~\PYTHIA~6 MC09 tune. The statistical uncertainties are
of the order of $\pm0.001$.}
\label{fig:etamatrix}
\end{figure}

To obtain an overall picture, and as a check on the experimental
approach, the FB multiplicity correlations using 7~TeV collisions were
calculated over a matrix of forward-backward pairs of $\eta$ intervals
of width $\delta\eta=0.5$ covering the range 0 to 2.5 in each
direction, for a \pTmin\ value of 100~MeV.  The results are given in
table~\ref{tab:matrix} and shown pictorially in
figure~\ref{fig:etamatrix}.  The main diagonal depicts the symmetric
FB correlations. Both for the data and for the chosen MC, the
correlations are seen to vary strongly with the separation between the
chosen intervals, and only weakly with the mean $\eta$ value for a
given separation, confirming that the symmetric forward-backward
configuration is a characteristic representation of the correlation at
a given separation in $\eta$.

\begin{figure}[!ht]
\centering
  \includegraphics[width=0.75\textwidth]{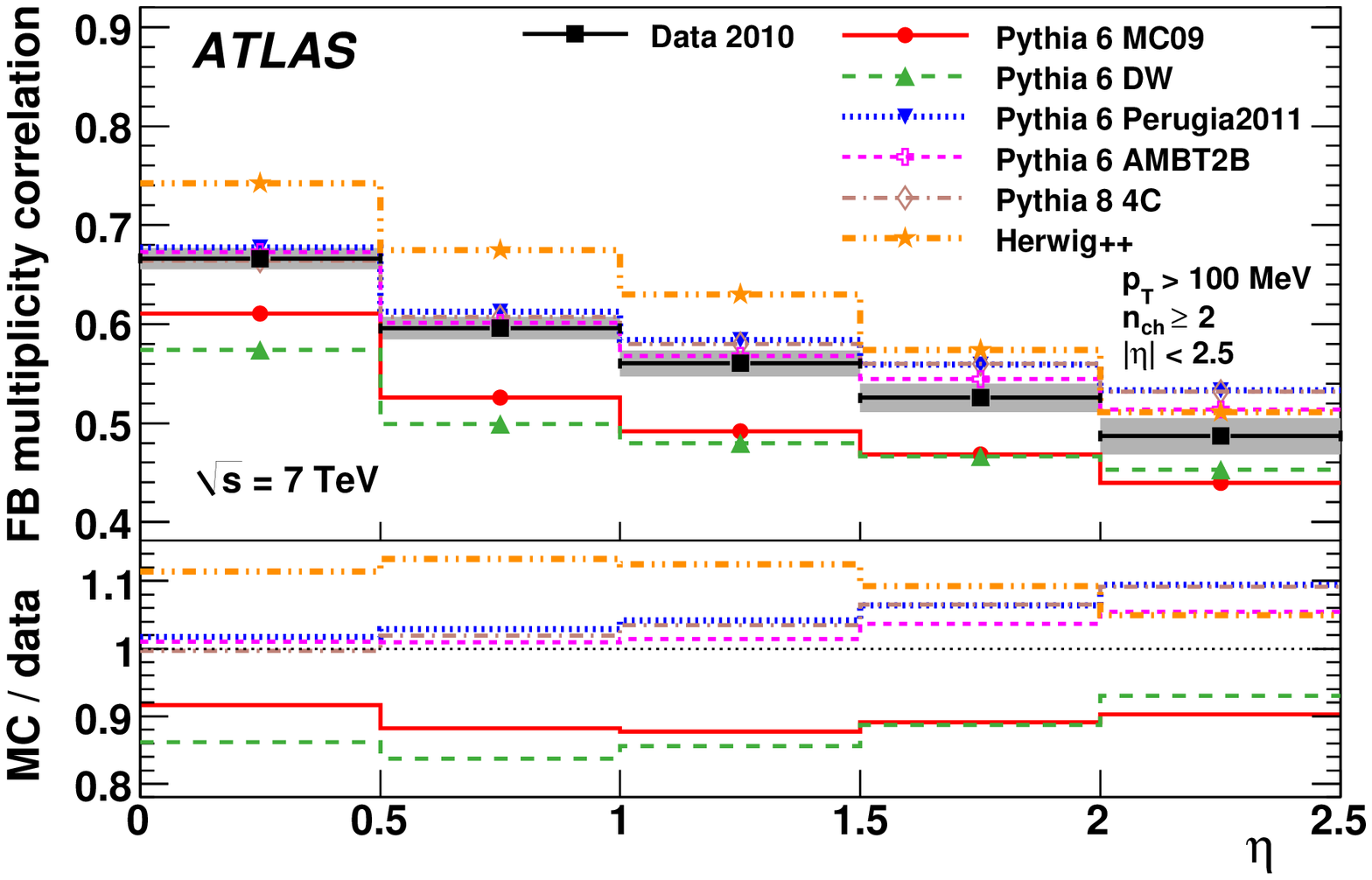}
  \\
  \includegraphics[width=0.75\textwidth]{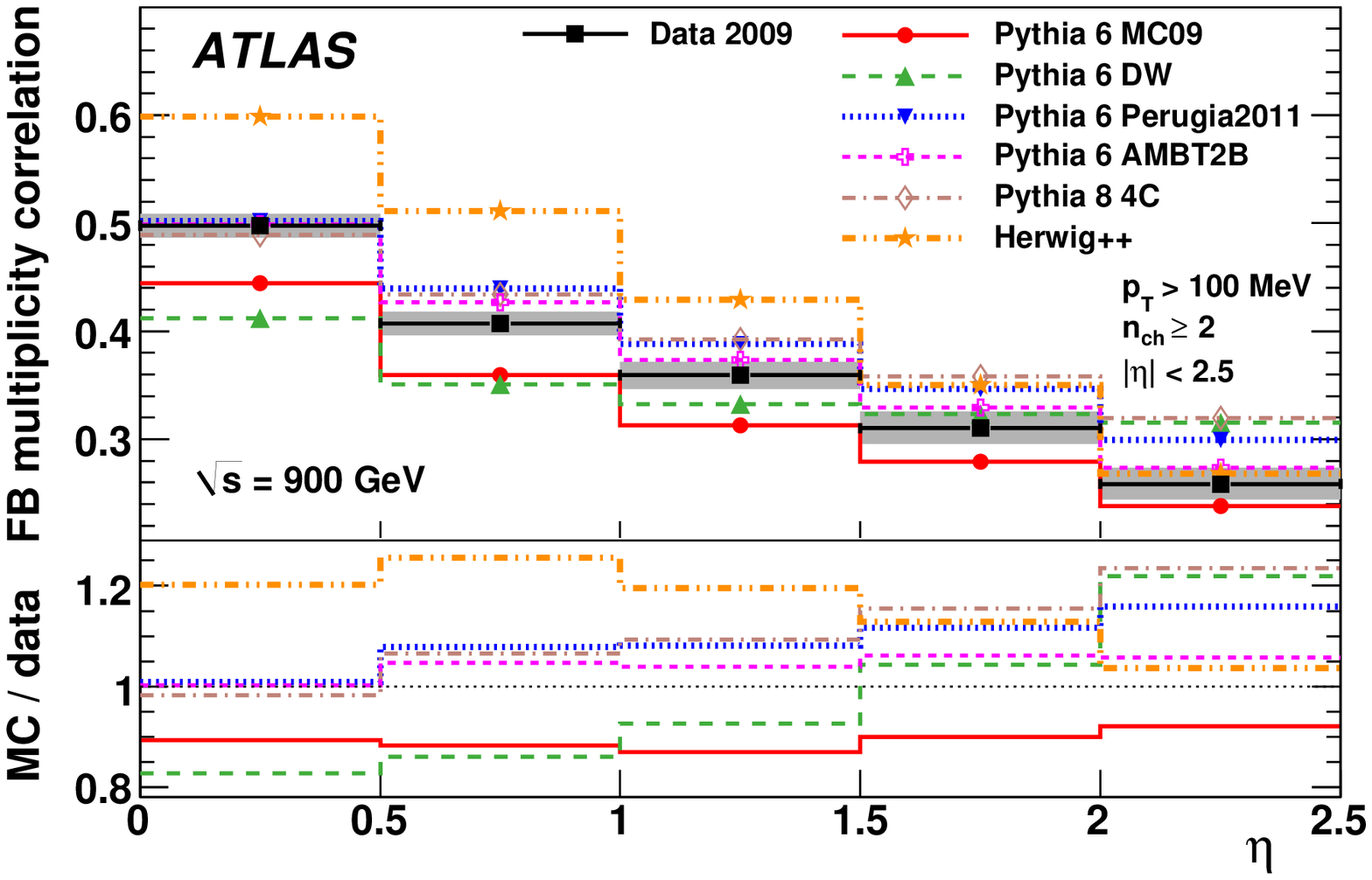}
  \\
  \includegraphics[width=0.75\textwidth]{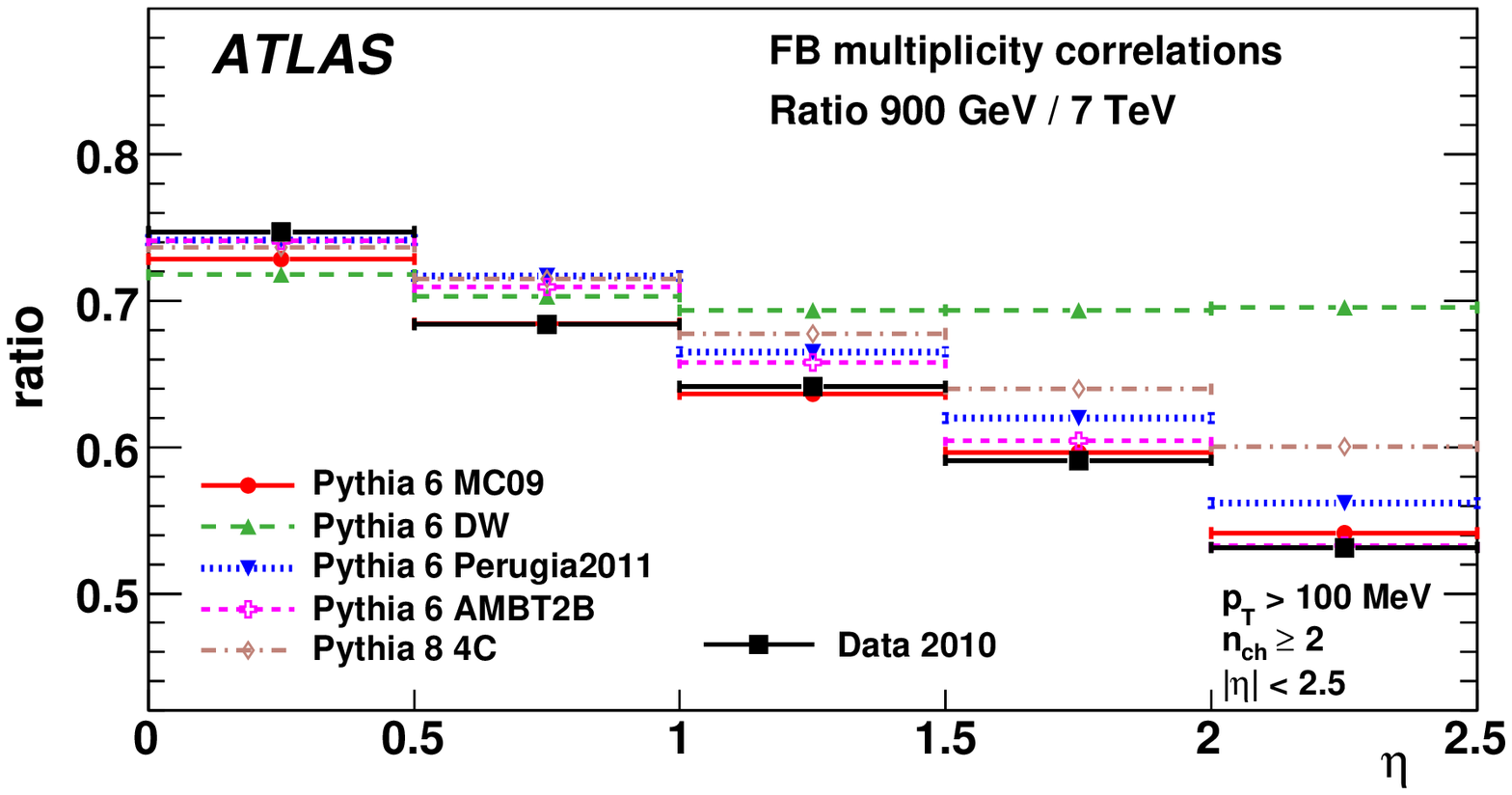}
\\[-0.916\textwidth](a)\\[0.445\textwidth](b)\\[0.345\textwidth](c)\\[0.0\textwidth]
\caption{Forward-backward multiplicity correlation in symmetrically
opposite $\eta$ intervals for events with at least two charged
particles with $p_{\rm{T}} \geq $ 100~MeV and $|\eta| < 2.5$. (a)~Data
at 7~TeV, compared with a selection of MC simulations. The systematic
uncertainties are indicated by a grey band; the statistical
uncertainties are too small to be visible on the figure. (b)~The same
at 900~GeV. (c)~Ratio of the 900~GeV results to the 7~TeV results.  }
\label{fig:all_bcorr}
\end{figure}

In the remainder of this section, the correlations between
symmetrically opposite FB pseudorapidity intervals of width
$\delta\eta=0.5$ are examined. A global correlation has also been
calculated, defined as the FB correlation taken over the entire
pseudorapidity range ($\delta\eta=2.5$).  The results are summarised
in table~\ref{tab:mult}.  Figure~\ref{fig:all_bcorr}(a) shows the
variation of the FB multiplicity correlation with $\eta$ found in the
7~TeV data, compared with various MC tunes.
Figure~\ref{fig:all_bcorr}(b) shows the same quantities measured at
900~GeV. The general shape of the data distributions is reasonably well
reproduced by the MCs, though there are discrepancies in detail. In
the case of MC09, the trend is to underestimate the correlation by
$\sim10\%$.  The correlation is generally similar for the DW and MC09
tunes at 7~TeV, but shows greater variation at 900~GeV where it rises
relative to the data at high $\eta$ values, implying an inaccuracy in
the modelled ratio of long-distance to short-distance
correlations. The \PYTHIA~8 tune performs well at low $\eta$ but
somewhat poorly at higher values.  The correlations given by AMBT2B
are the most consistent with the data; AMBT2B was tuned to these event
samples using different variables. At both energies \HERWIG++ gives
correlations consistently higher than the data.

Figure~\ref{fig:all_bcorr}(c) shows the ratio of the FB multiplicity
correlation distributions at 900~GeV and 7~TeV, compared with the
results from the MC tunes. The multiplicity correlations are
substantially lower at 900~GeV than at 7~TeV and the relative
difference is greater for the larger pseudorapidity bins. This is a new
experimental result which can be interpreted in general terms as
indicating that, while the short-range correlation is similar at the
two energies, the long-range correlation becomes considerably higher
as the energy increases. This trend is reproduced in most of the 
MC predictions, although not so well by DW. The \HERWIG++ ratio is
not included since different tunes are used at the two energies.

\begin{figure}[!ht]
\centering
\includegraphics[width=0.75\textwidth]{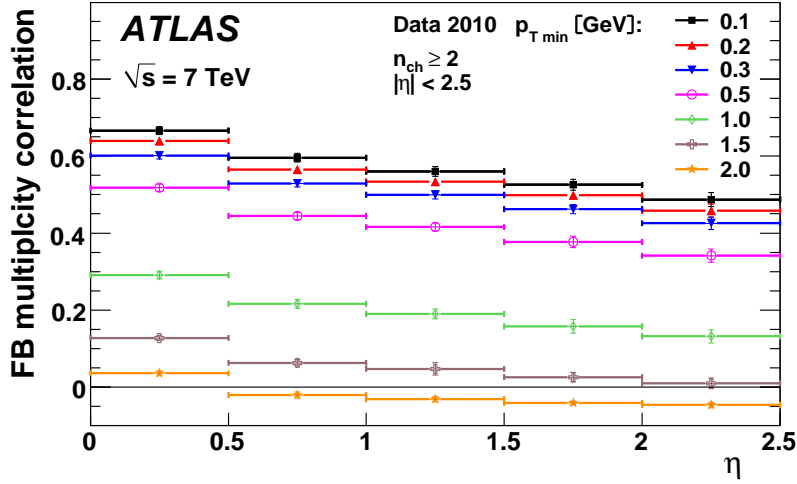}
\includegraphics[width=0.75\textwidth]{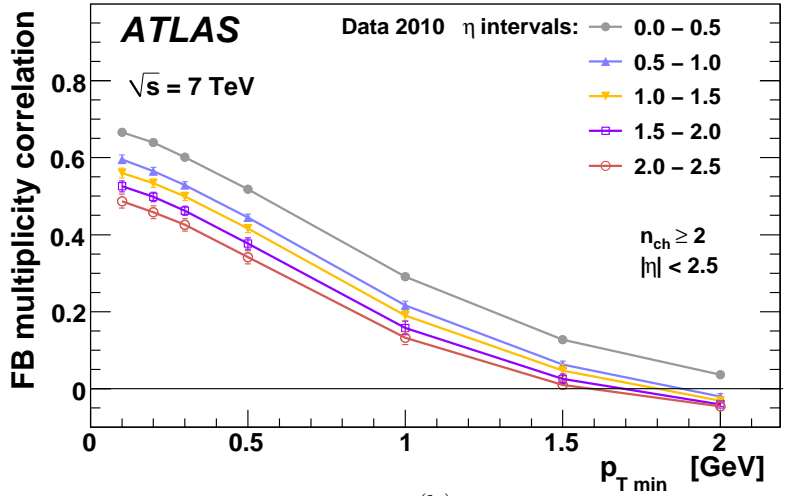}
\\[-0.5\textwidth](a)\\[0.45\textwidth](b)\\[0.0\textwidth]
\caption
{ Forward-backward multiplicity correlations: (a)~as a function of
$\eta$ for different values of $p_{\rm{ T min}}$; (b)~as a function of
$p_{\rm {T min}}$ for different values of $\eta$, with lines drawn
to guide the eye. The plotted systematic uncertainties 
are smaller than the symbol in many cases; the statistical
uncertainties are too small to be visible on the figure.  }
\label{fig:manypt}
\end{figure}

The minimum value, \pTmin, of the transverse momentum of the selected
charged particles was varied for the 7~TeV data. The resulting
multiplicity correlations are shown in table~\ref{tab:manypt} and
illustrated in figure~\ref{fig:manypt}. As expected, the correlations
fall rapidly as \pTmin\ increases above a few hundred~MeV, a feature
also seen in the MC models (not shown). In the context of the Lund
string model~\cite{ref:strings1}, at low transverse momentum values,
this may be seen as a general tendency for a partonic string to
fragment in a uniform way all along its length. There is also a
possible influence of MPI effects.  At higher transverse momentum,
however, particles are more likely to be associated with jets,
and there is no strong correlation between a given jet and another jet
at any particular value of $\eta$. The tendency to have negative
correlation values at higher \pT\ is a manifestation of the tendency
of high-\pT\ particles to be collimated into jets.

\begin{figure}[!ht]
\centering
  \includegraphics[width=0.75\textwidth]{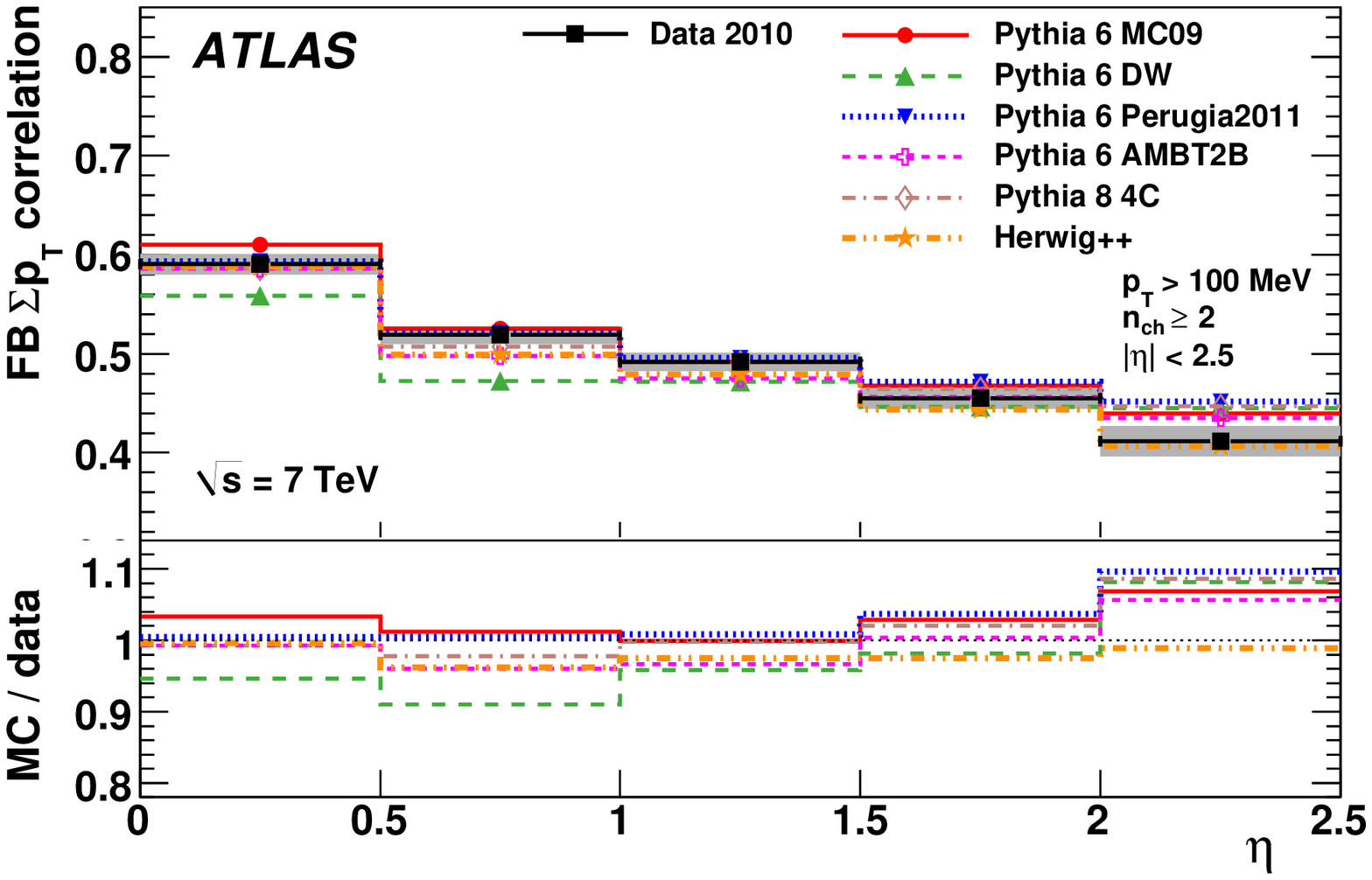}
  \\
  \includegraphics[width=0.75\textwidth]{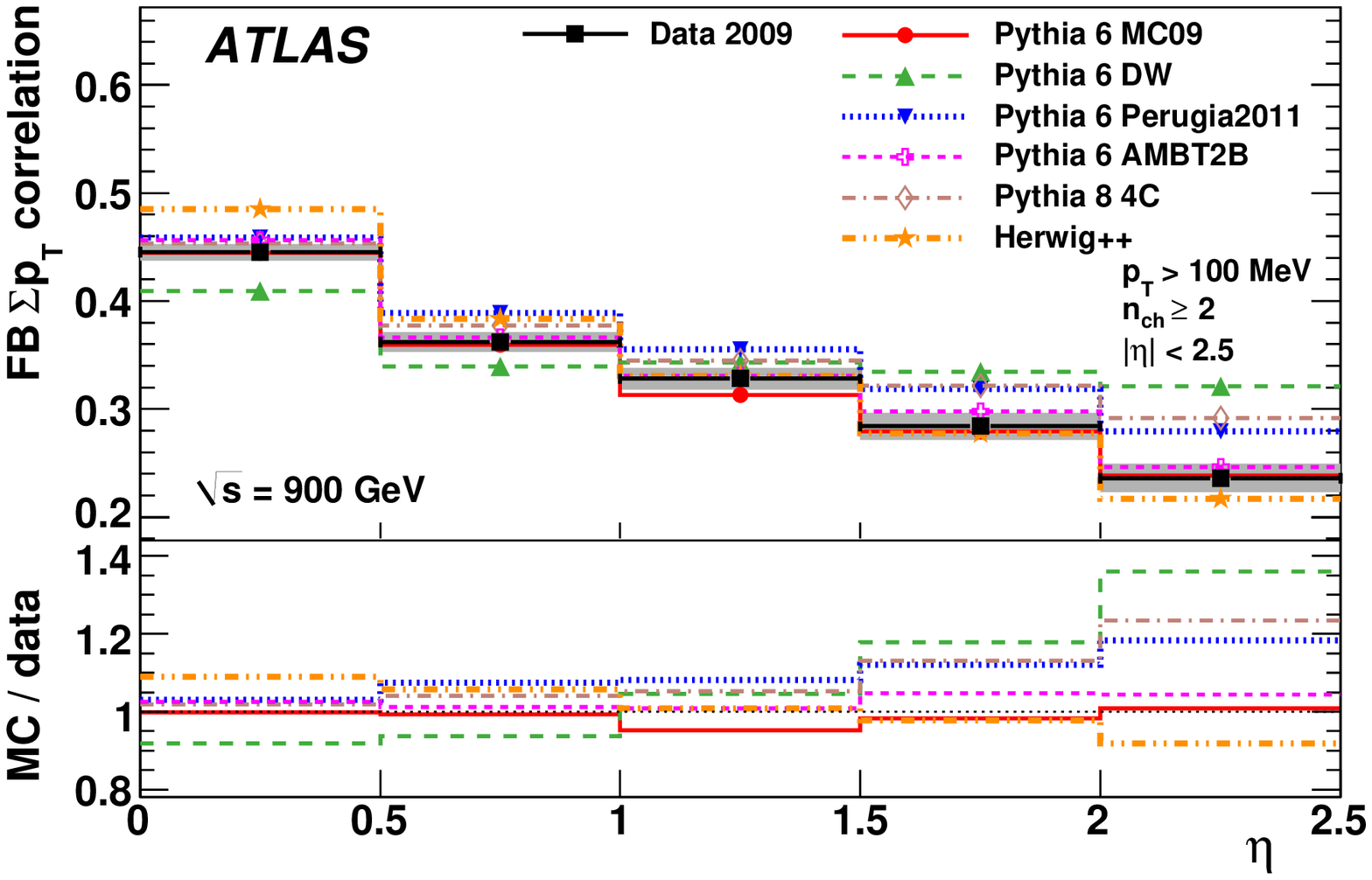}
  \\
  \includegraphics[width=0.75\textwidth]{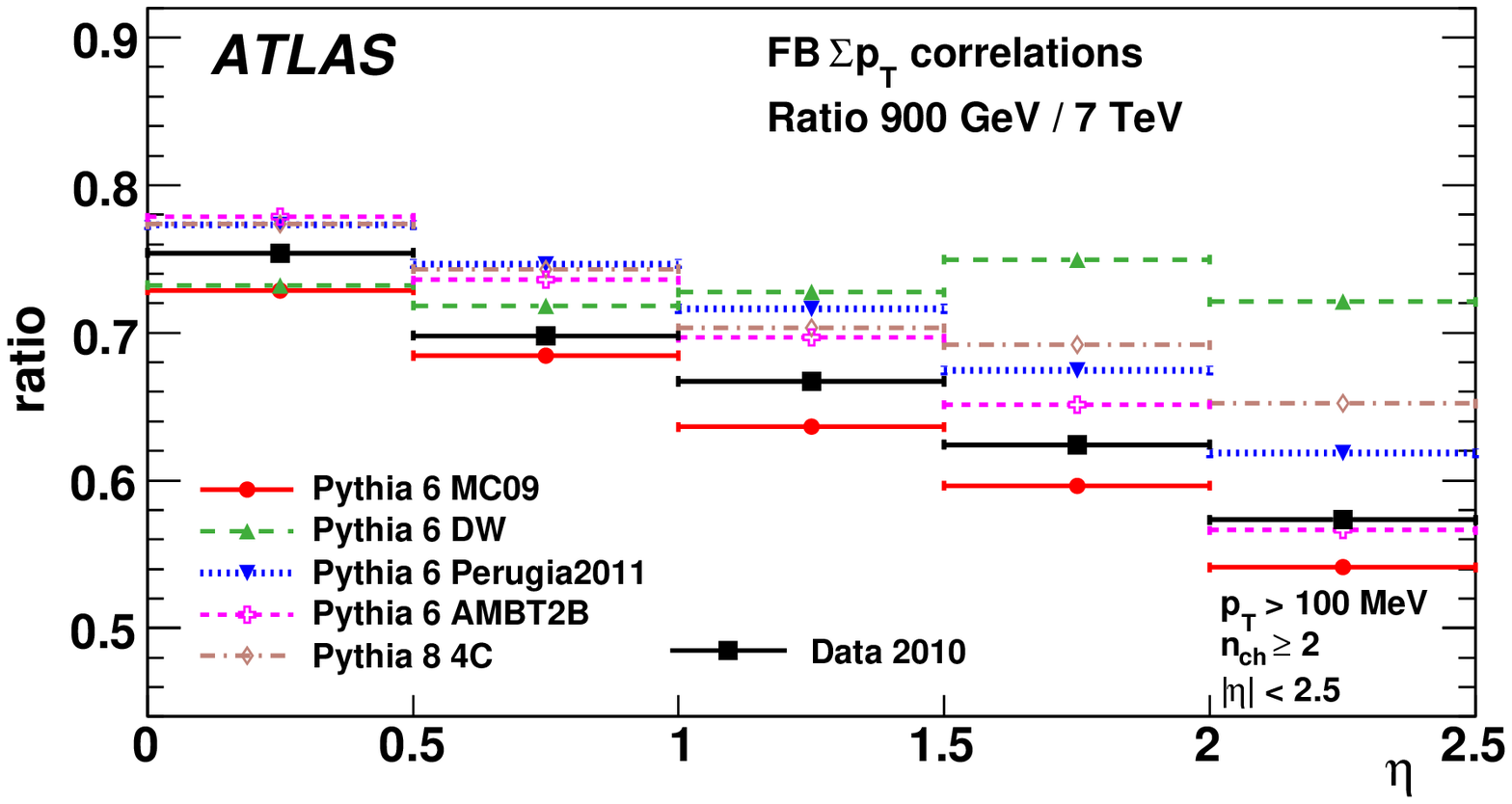}
\\[-0.92\textwidth](a)\\[0.445\textwidth](b)\\[0.355\textwidth](c)\\[0.0\textwidth]
\caption{Forward-backward $\Sigma$\pT\ correlation in symmetrically
opposite $\eta$ intervals for events with at least two charged
particles with $p_{\rm{T}} \geq $ 100~MeV and $|\eta| < 2.5$. (a)~Data
at 7~TeV, compared with a selection of MC simulations. The systematic
uncertainties are indicated by a grey band; the statistical
uncertainties are too small to be visible on the figure. (b)~The same
at 900~GeV. (c)~Ratio of the 900~GeV results to the 7~TeV results.  }
\label{fig:sumpt}
\end{figure}

The FB correlations between the summed \pT\ of charged particles are shown
in figure~\ref{fig:sumpt} and listed in table~\ref{tab:sumpt}.  In
general, the features are similar to those of the multiplicity
correlations.  As in the case of the multiplicities, the momentum
FB correlations are higher at 7~TeV than at 900~GeV.

\subsection{Azimuthal distributions}

\begin{figure}[!ht]
\centering
  \includegraphics[width=0.48\textwidth]{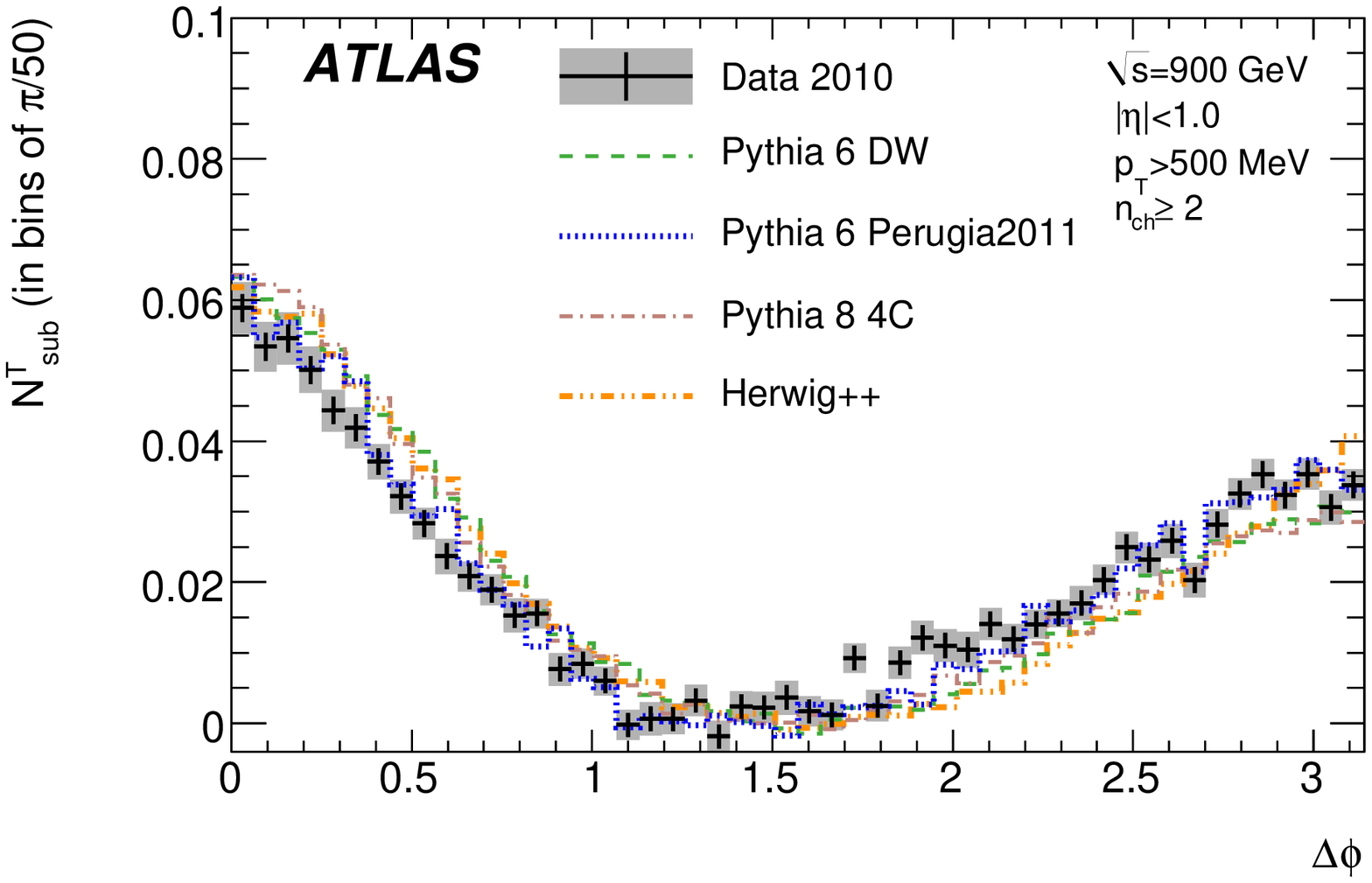}
  \includegraphics[width=0.48\textwidth]{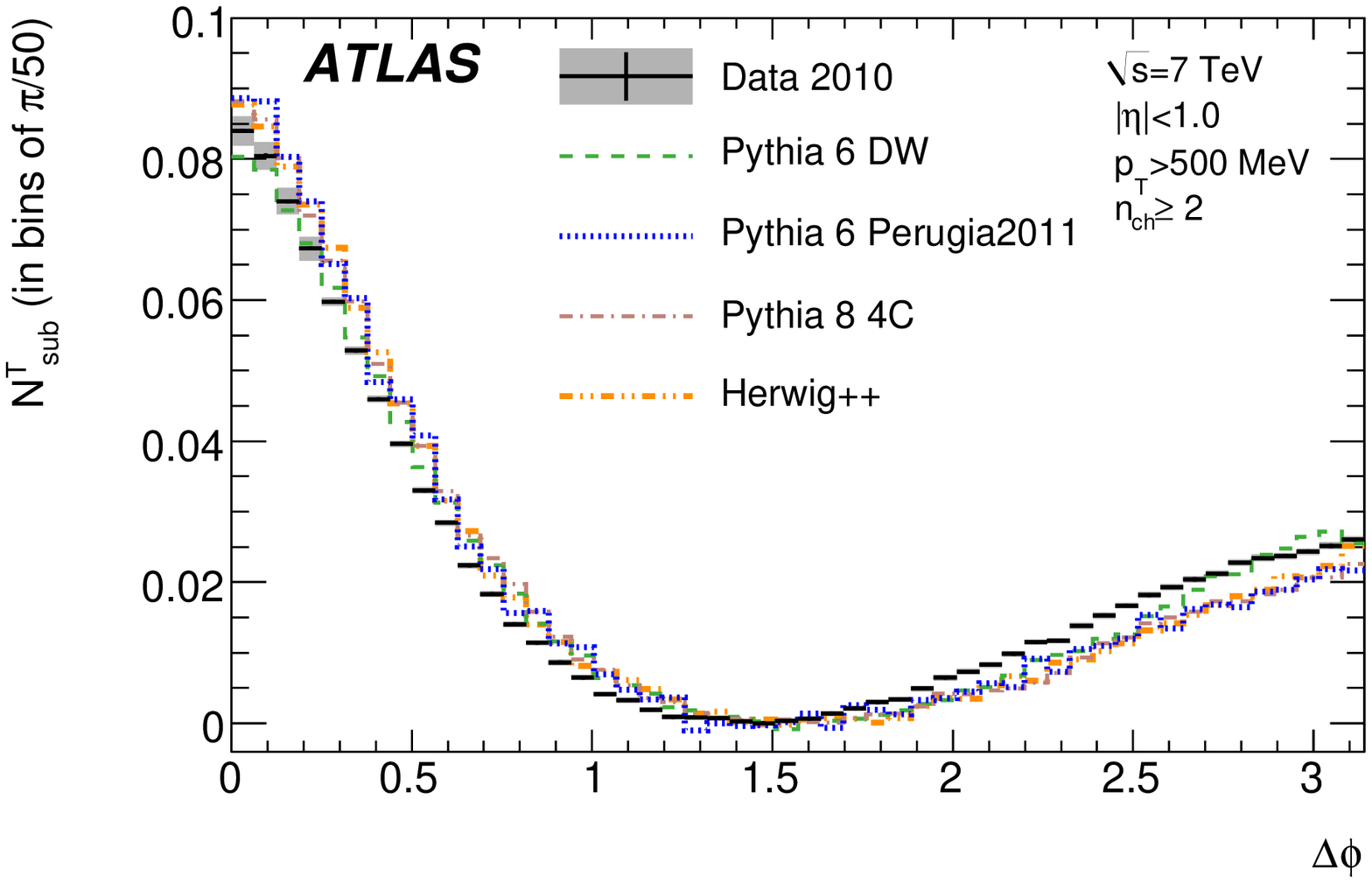}
\\[-1.9ex]\hspace*{3.5ex}(a)\hspace*{0.45\textwidth}(b)\\[1.5ex]
  \includegraphics[width=0.48\textwidth]{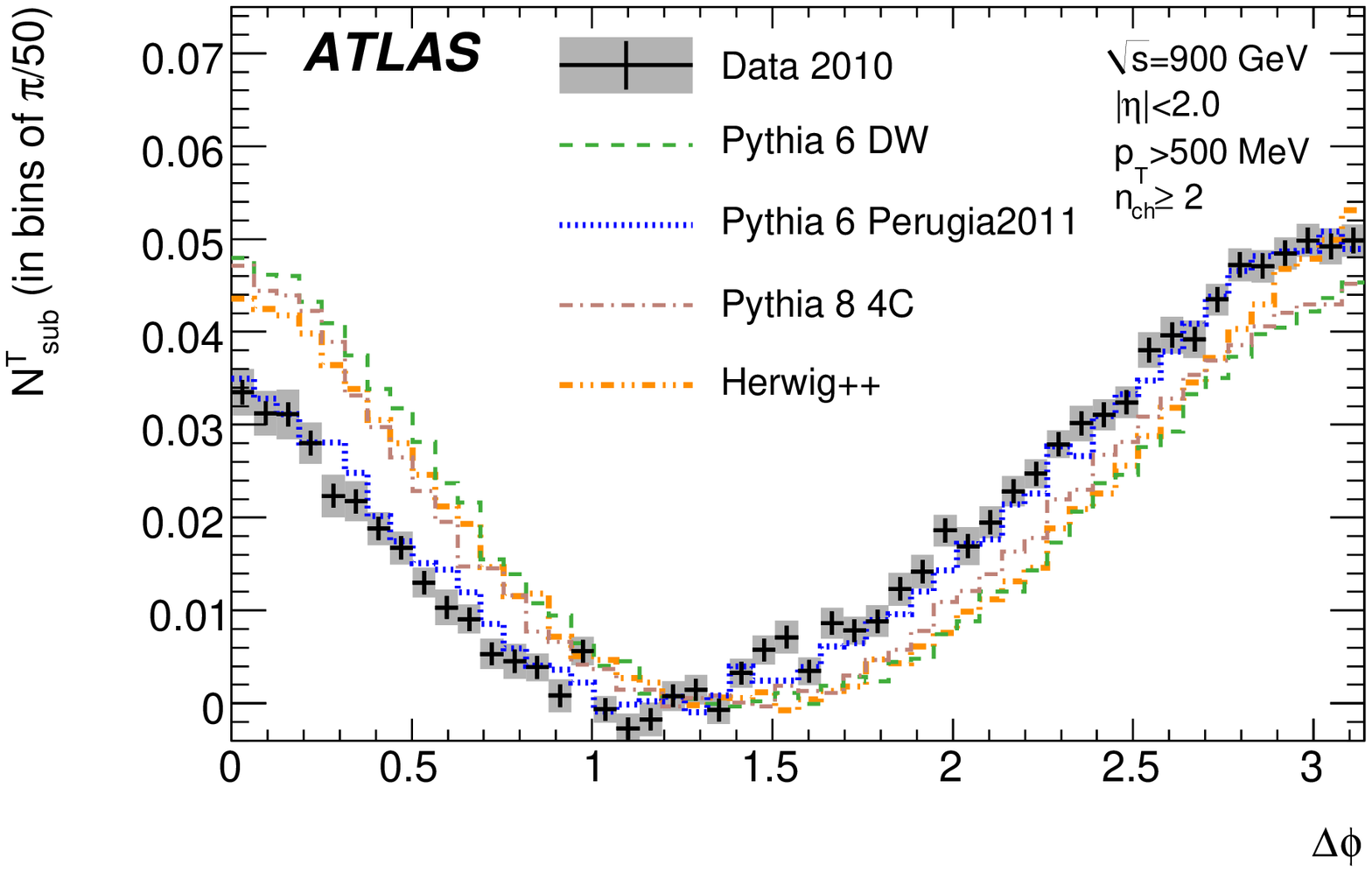}
  \includegraphics[width=0.48\textwidth]{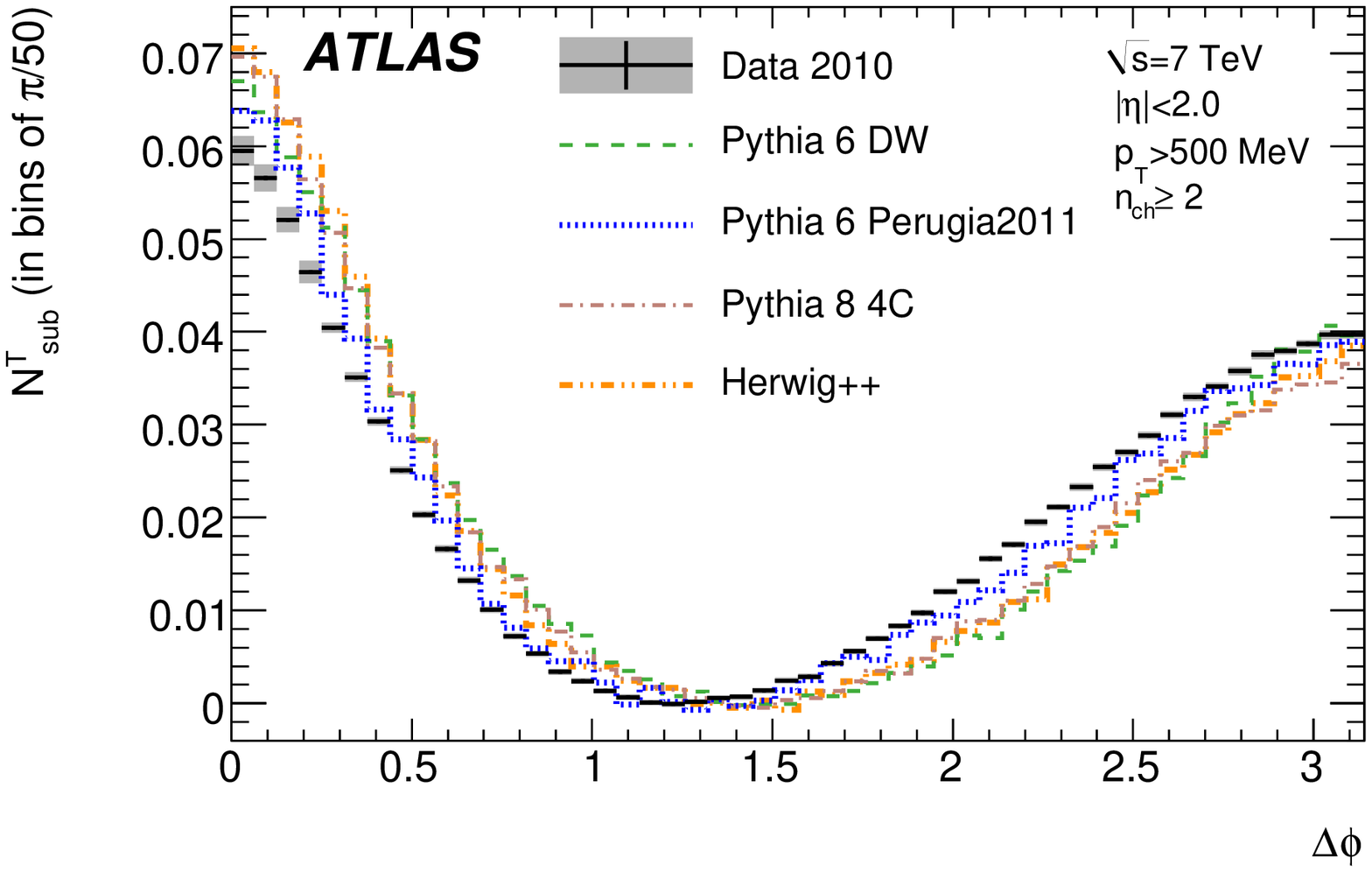}
\\[-1.9ex]\hspace*{3.5ex}(c)\hspace*{0.45\textwidth}(d)\\[1.5ex]
  \includegraphics[width=0.48\textwidth]{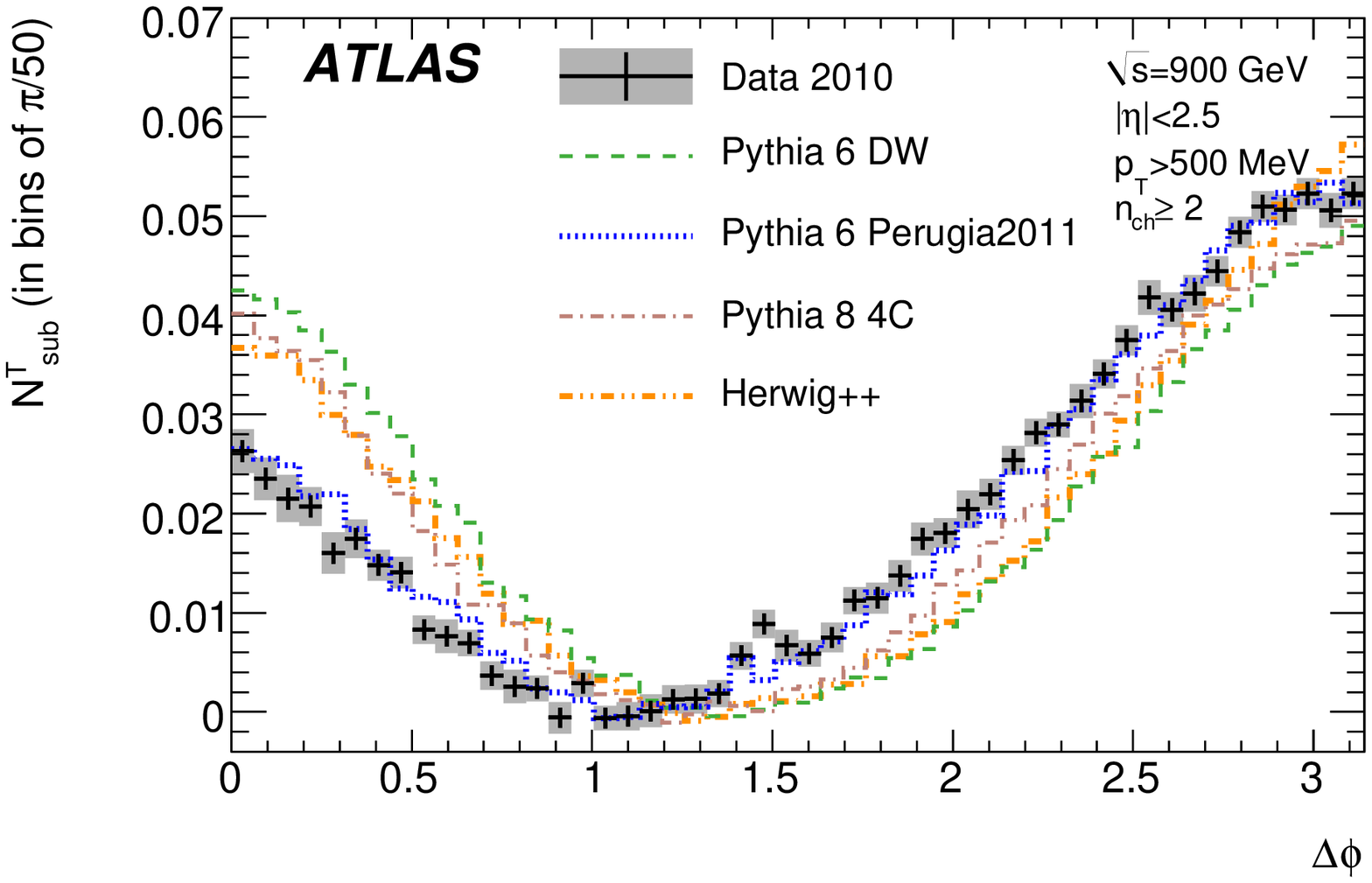}
  \includegraphics[width=0.48\textwidth]{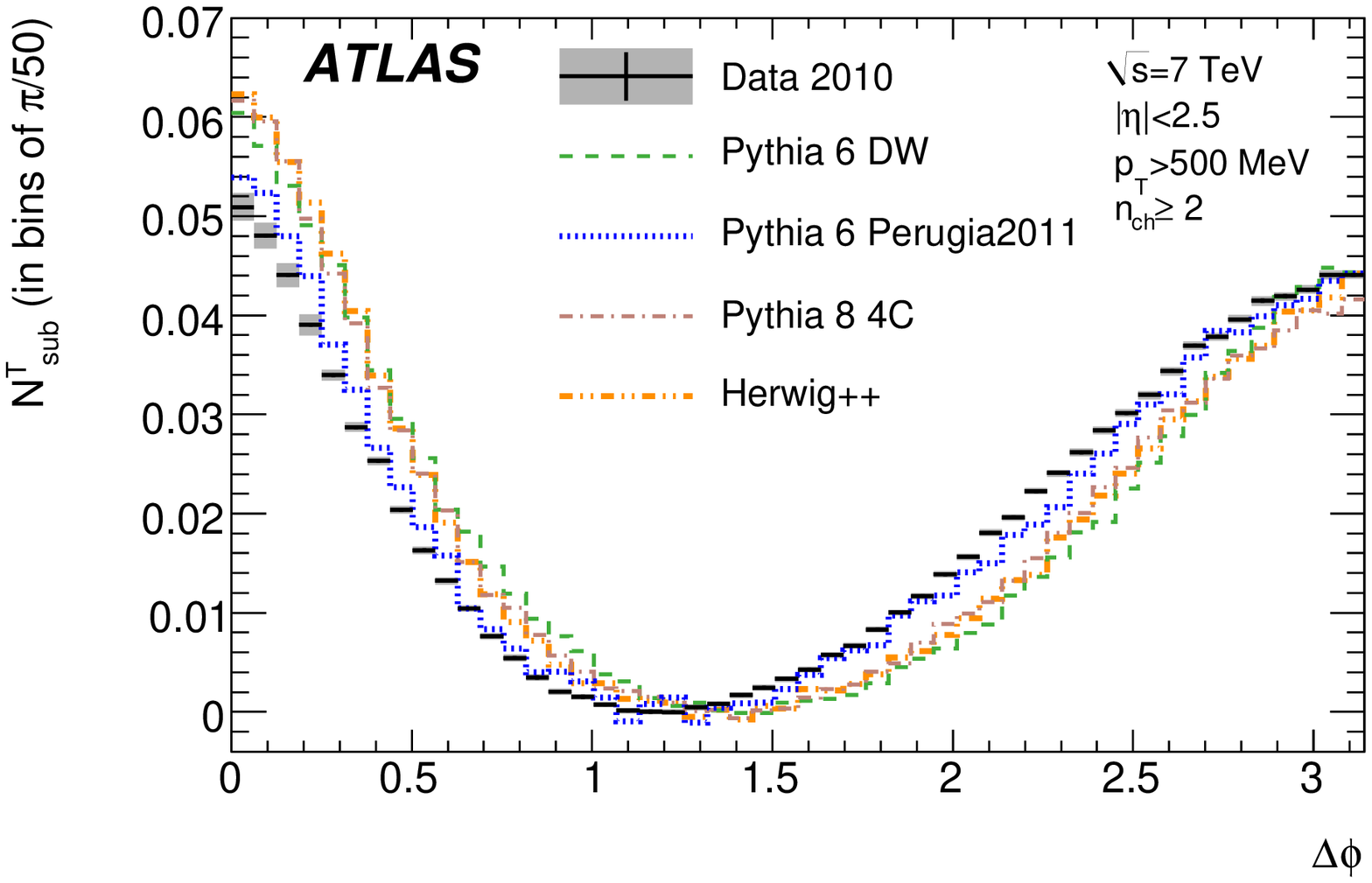}
\\[-1.9ex]\hspace*{3.5ex}(e)\hspace*{0.45\textwidth}(f)\\[1.5ex]
\caption{ Subtracted and normalised $\Delta\phi$ distributions
($N^T_{sub}$) for data taken at $\sqrt{s} = 900$~GeV (a, c, e) and 7
TeV (b, d, f). Three different pseudorapidity regions are shown:
$|\eta| < 1.0,$ $|\eta| < 2.0,$  $|\eta| < 2.5.$  The data (black
symbols) are compared with the predictions of different MC tunes
(see text for more details). The vertical bars on the data points
denote the statistical uncertainty, while the shaded areas denote the
total uncertainty. The leading particle in each event is omitted from the plot.  }
\label{fig:delphisub}
\end{figure}

\begin{figure}[!ht]
\centering
  \includegraphics[width=0.48\textwidth]{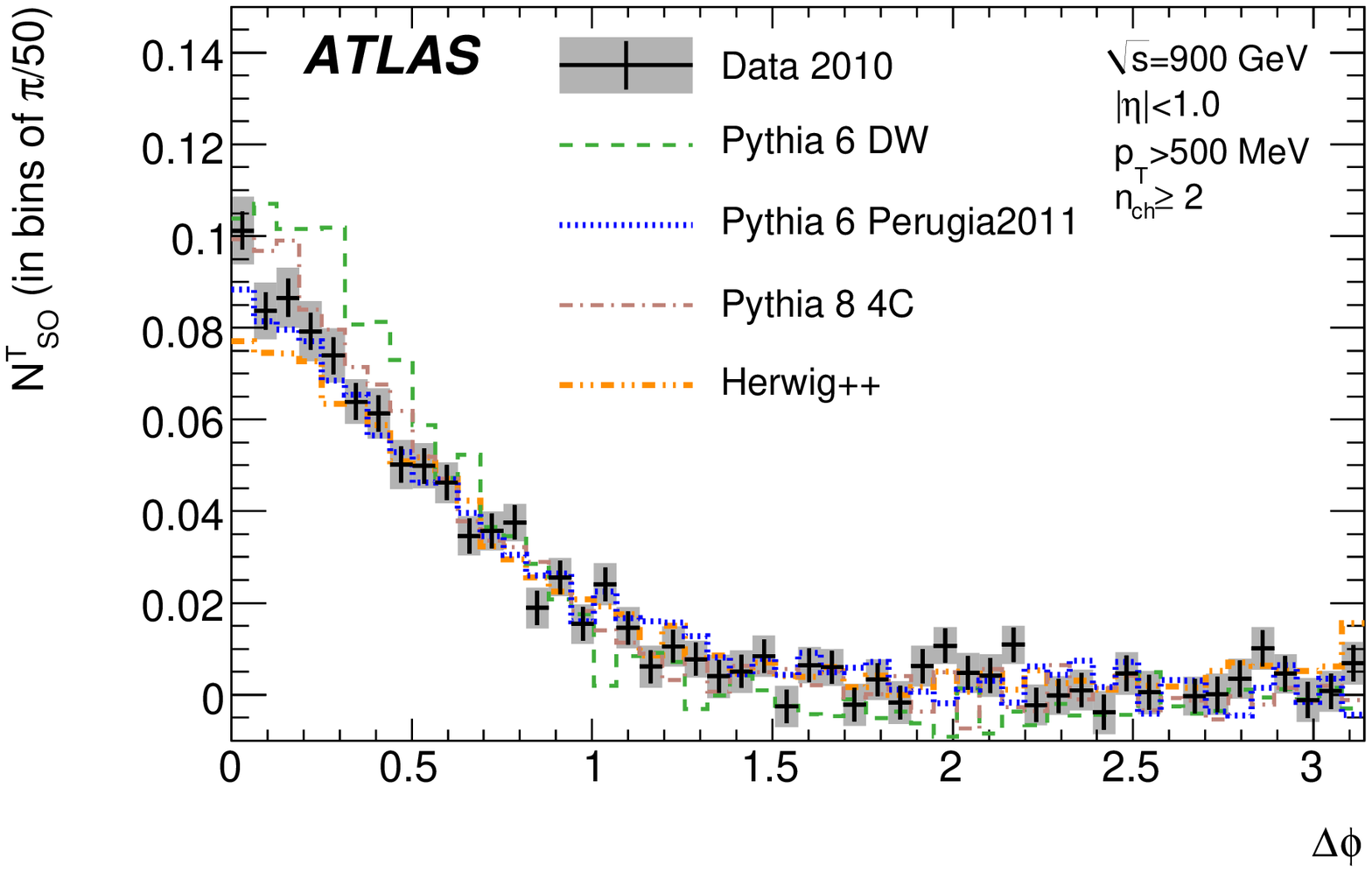}
  \includegraphics[width=0.48\textwidth]{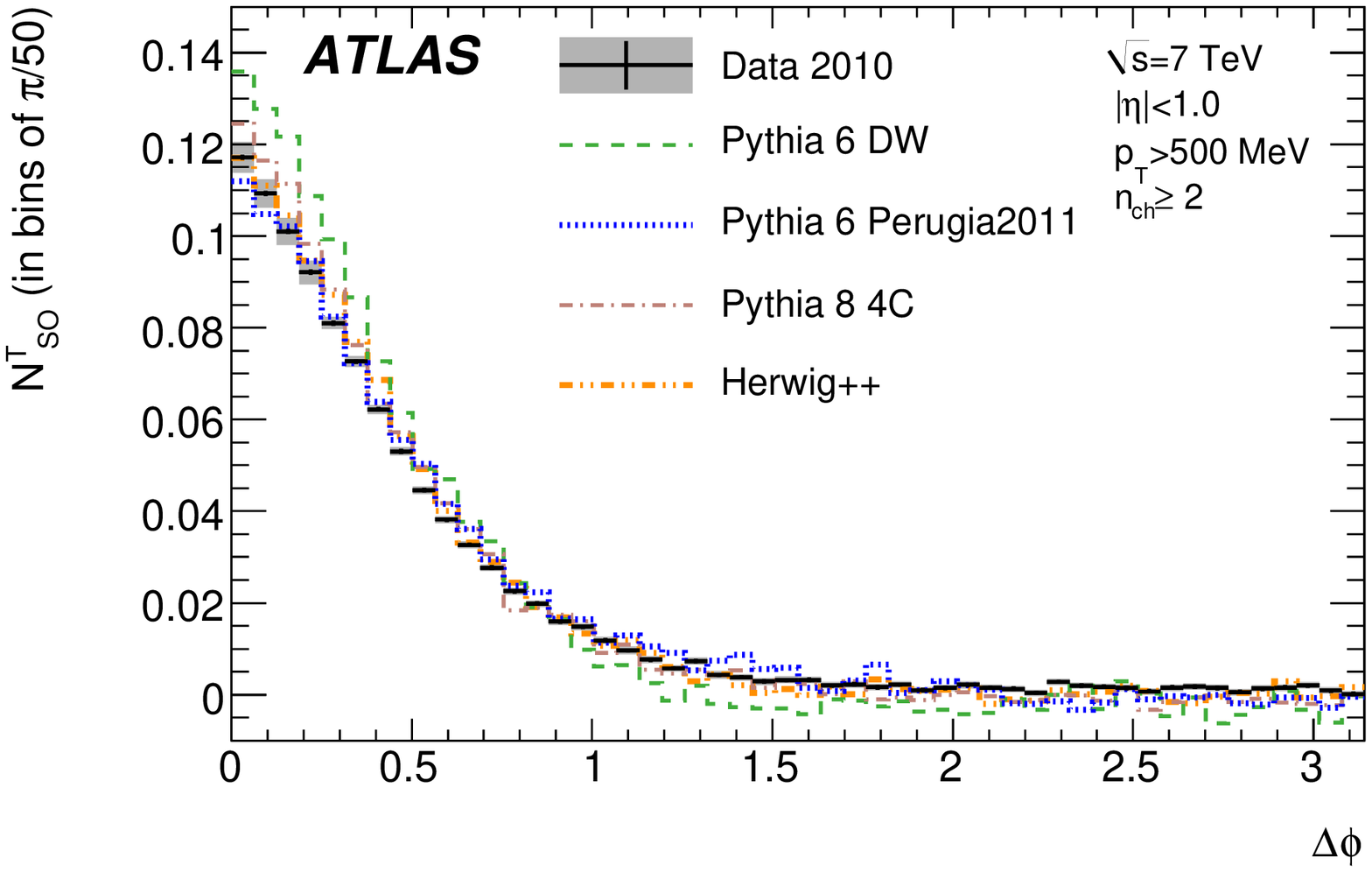}
\\[-1.9ex]\hspace*{3.5ex}(a)\hspace*{0.45\textwidth}(b)\\[1.5ex]
  \includegraphics[width=0.48\textwidth]{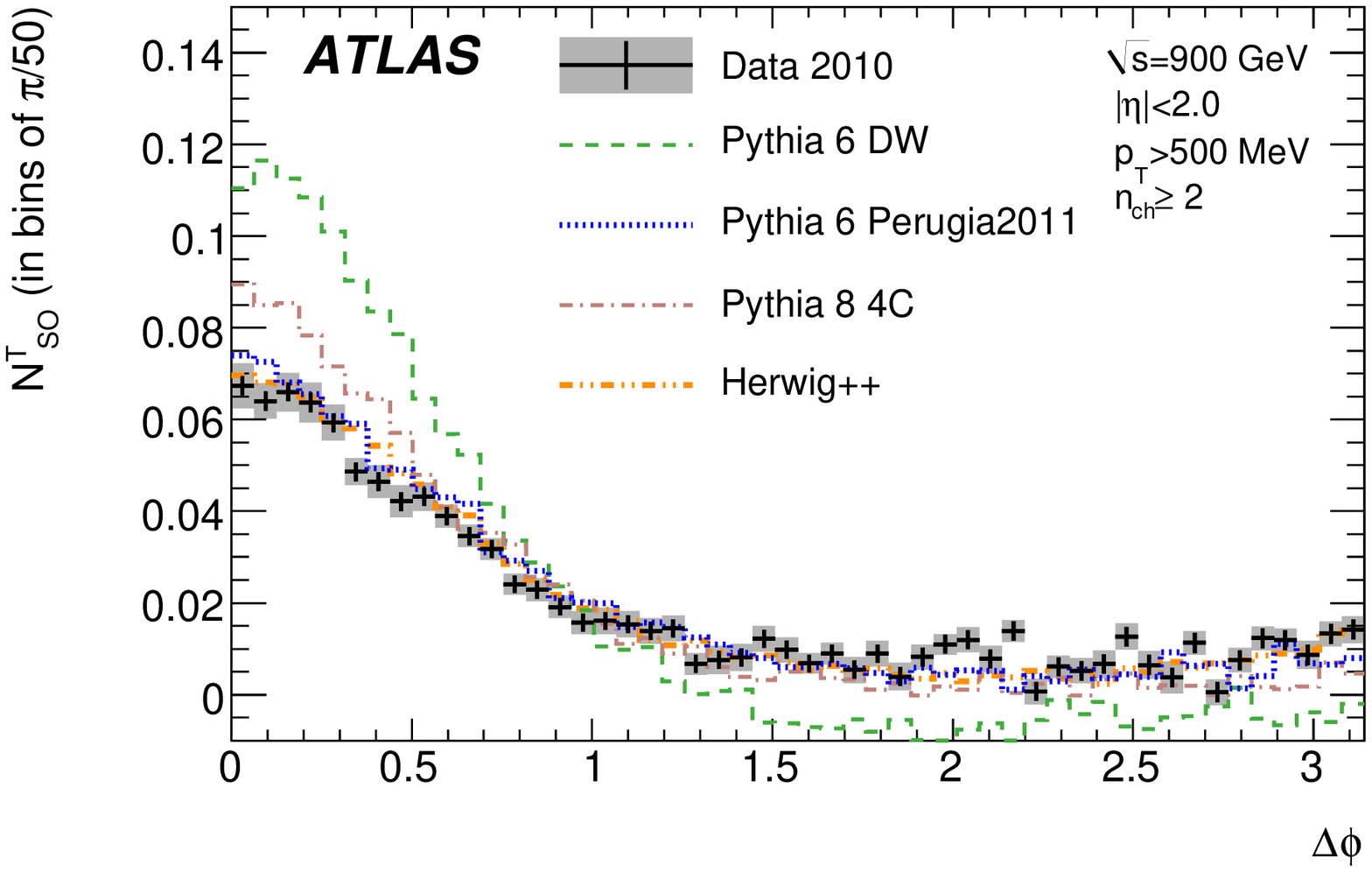}
  \includegraphics[width=0.48\textwidth]{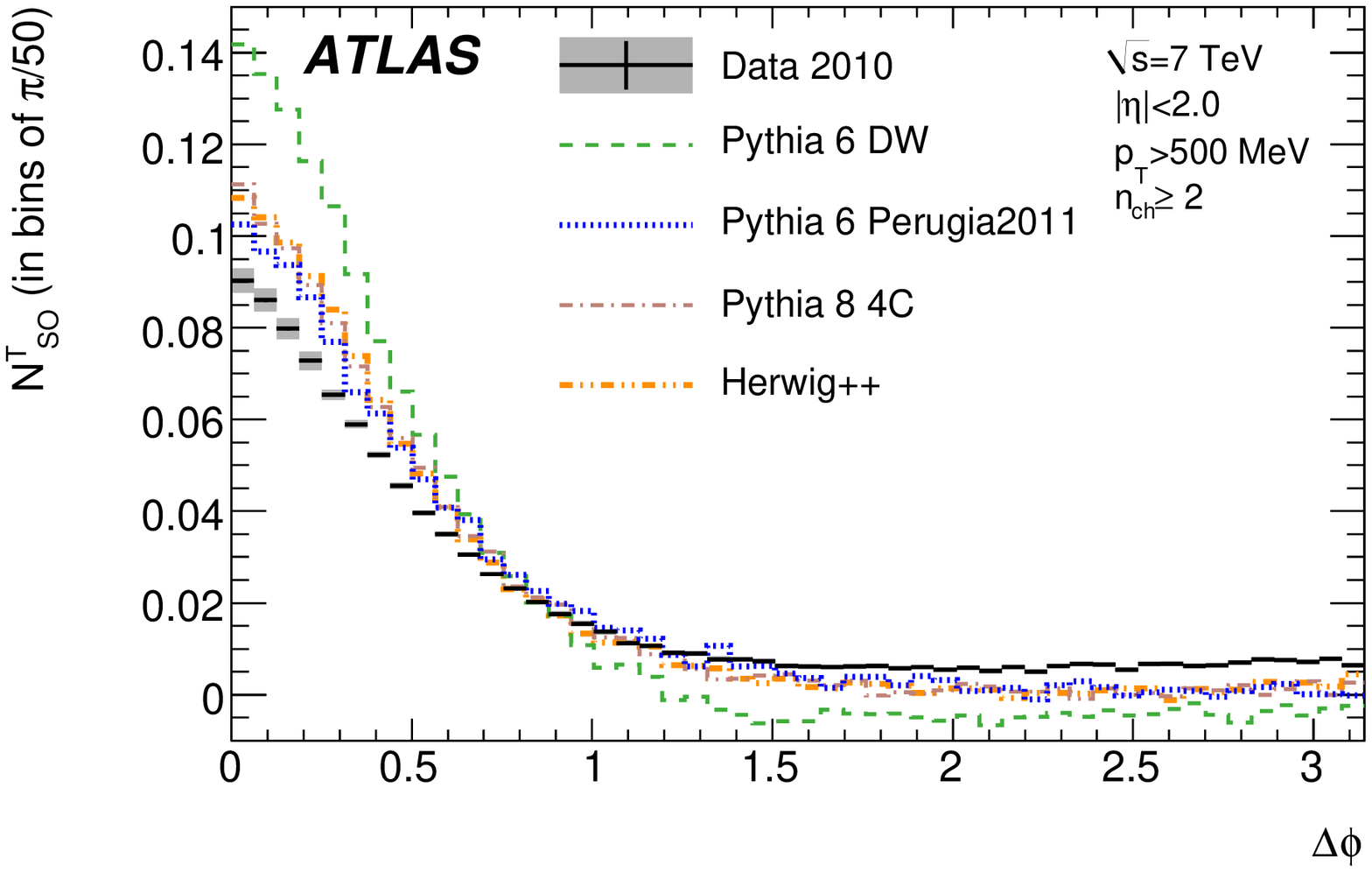}
\\[-1.9ex]\hspace*{3.5ex}(c)\hspace*{0.45\textwidth}(d)\\[1.5ex]
  \includegraphics[width=0.48\textwidth]{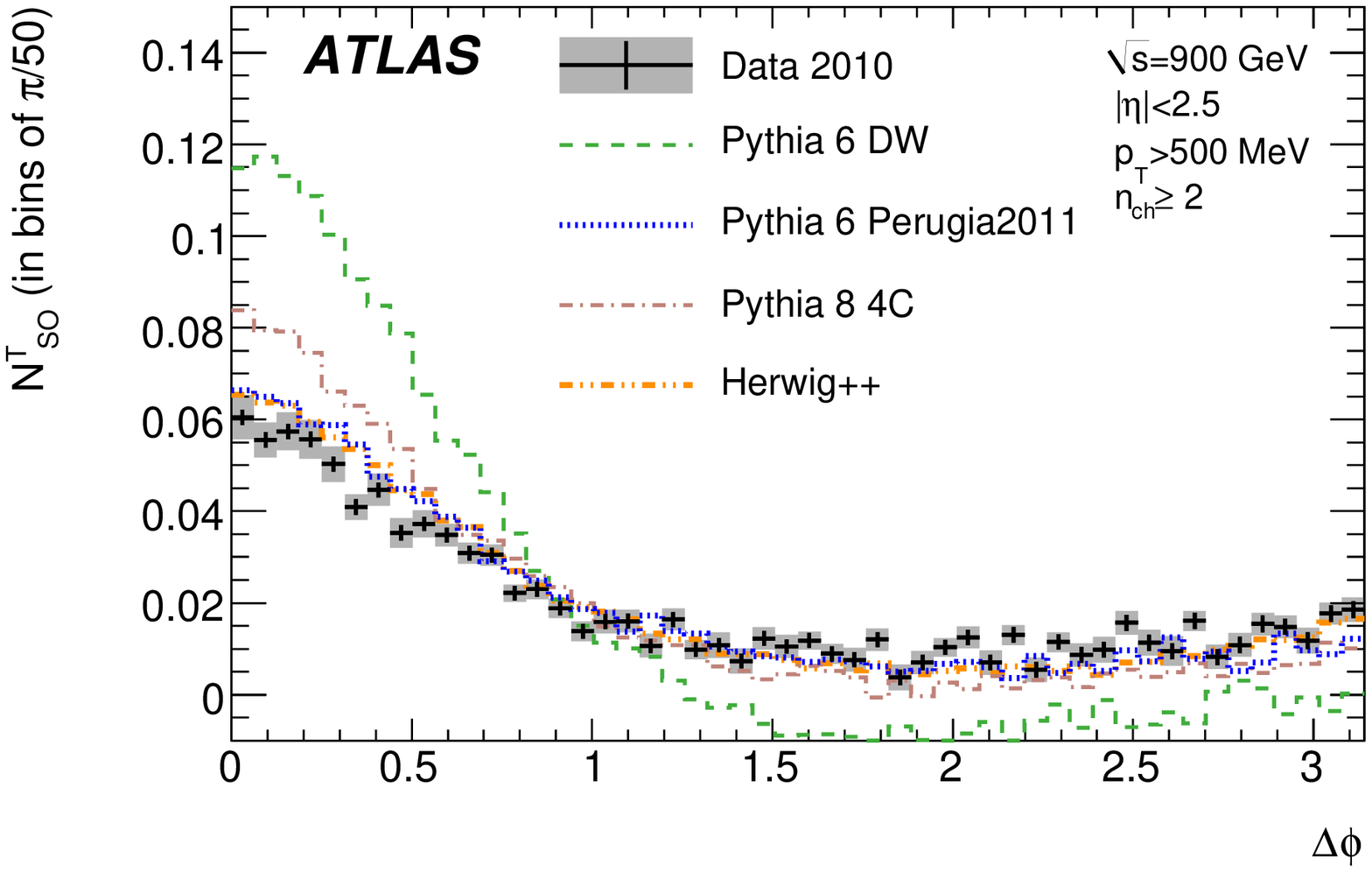}
  \includegraphics[width=0.48\textwidth]{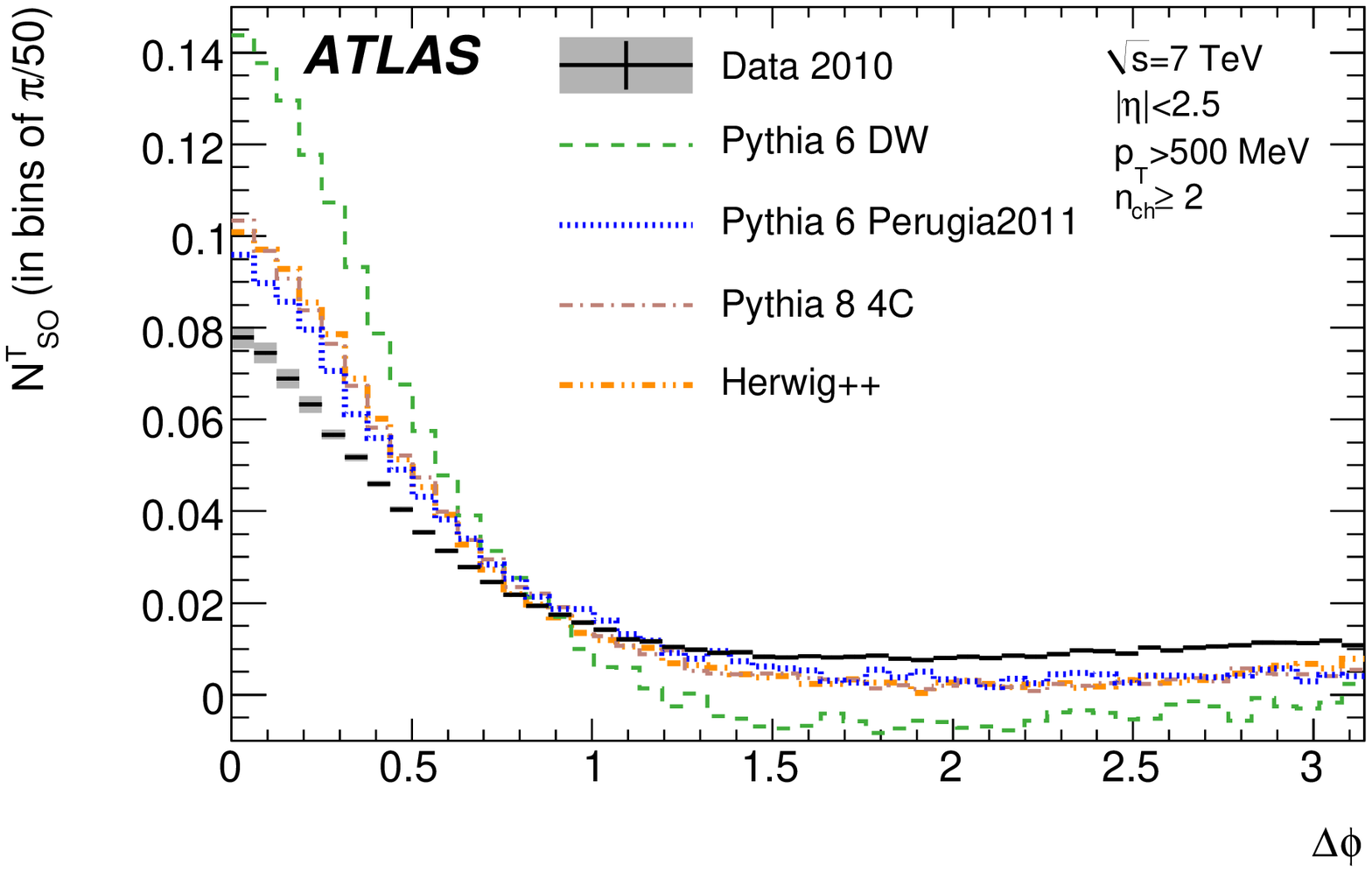}
\\[-1.9ex]\hspace*{3.5ex}(e)\hspace*{0.45\textwidth}(f)\\[1.5ex]
\caption{Normalised ``same hemisphere minus opposite'' distributions
($N^T_{SO}$) for data taken at $\sqrt{s} = 900 $~GeV (a, c, e) and 7
TeV (b, d, f). Three different pseudorapidity regions are shown:
$|\eta| <$ 1.0, $|\eta| <$ 2.0, $|\eta| <$ 2.5.  The data (black
symbols) are compared with the predictions of different MC tunes
(see text for more details). The vertical bars on the data points
denote the statistical uncertainty, while the shaded areas denote the
total uncertainty. The leading particle in each event is omitted from
the plot.  }
\label{fig:delphismo}
\end{figure}

Distributions of the subtracted and the ``same hemisphere minus
opposite'' charged-particle distributions in azimuthal angle,
$N^T_{sub}$ and $N^T_{SO}$, are shown in figures \ref{fig:delphisub}
and \ref{fig:delphismo}, respectively, for $\sqrt{s} =$ 900~GeV and
7~TeV, and compared with a selection of the different MC tunes. The
leading particle in each event is omitted from the plots. One notes in
figure~\ref{fig:delphisub} the increasing prominence of the
``opposite'' jet peak in the distributions as the $|\eta|$ interval
increases, as more pairs of jets, separated in $\eta$, become included
within the acceptance.  The shapes show a strong energy dependence.

     In general, the MC tunes and models used for comparison with data
 provide a reasonable description in the restricted rapidity
 range $|\eta| < 1$, but the agreement becomes increasingly poor as
 the $|\eta|$ range is increased to 2 and 2.5.  For the subtracted
 distributions the Perugia2011 tune is the most successful, giving a
 good representation of the distribution at 900 GeV for all
 three $|\eta|$ ranges.  At 7 TeV it is poorer for $|\eta| < 1,$ but
 similar to the other tunes and models presented, and it improves as
 the coverage in $|\eta|$ increases.  The other three models are very
 similar at 900 GeV, where they all give an equally poor description
 of the data in the broader $|\eta|$ ranges.  At 7 TeV, DW
 describes the data well for $|\eta| < 1,$ but is a poorer match in 
 the higher $|\eta|$ ranges.  \HERWIG++ and \PYTHIA~8 are similar at
 all $|\eta|$, giving a somewhat poor description that resembles
 Perugia2011 at $|\eta|<1$ and DW at higher $|\eta|$.  In all
 cases, the discrepancies of the tunes and models tend towards
 over-enhancing the same-side peak relative to the opposite-side peak.

    The same-minus-opposite variable is described well for $|\eta| <
 1$ at both energies by \PYTHIA~8, Perugia2011 and \HERWIG++, with
 regard to both the leading-particle ``jet peak" and the opposing
 pedestal.  However DW exaggerates the sharpness of the peak.
 As the accepted range in $|\eta|$ increases, so that the averaged
 effects of more widely-separated jet pairs become included in the
 distributions, the disagreements become more pronounced.  At both
 energies DW's exaggeration of the peak increases
 dramatically with $|\eta|$, while \PYTHIA~8 gives a progressively
 worse description of the peak, also overestimating its size.
 The \HERWIG++ and Perugia2011 models are very similar in this
 variable, and describe the data well at 900 GeV but less so at 7 TeV,
 where they also resemble \PYTHIA~8.  Again, the discrepancies tend
 towards an over-enhancement of the same-side ``jet peak'', which becomes a
 particularly strong effect in the case of DW.


\begin{sidewaystable}[!ht]
{
\begin{center}
\begin{tabular}{|c|c|c|c|c|c|}
\hline
Forward $\eta$ interval\vstrut & 0.0 -- 0.5 & 0.5 -- 1.0 & 1.0 -- 1.5 & 1.5 -- 2.0 & 2.0 -- 2.5 \\
Backward $\eta$ interval & & & & & \\
\hline
0.0 -- 0.5 & 0.666 & 0.624 & 0.592 & 0.566 & 0.540 \\
           & $\pm0.011$ & $\pm0.011$ & $\pm0.011$ &  $\pm0.012$ & $\pm0.013$ \\
\hline
0.5 -- 1.0 & 0.624 & 0.596 & 0.574 & 0.553 & 0.530 \\
           & $\pm0.011$ &  $\pm0.011$ &  $\pm0.012$ &  $\pm0.013$ &$\pm0.014$ \\
\hline
1.0 -- 1.5 & 0.594 & 0.576 & 0.560 & 0.540 & 0.518 \\
           &  $\pm0.011$ &  $\pm0.012$ & $\pm0.013$ &$\pm0.014$ & $\pm0.014$ \\
\hline 
1.5 -- 2.0 & 0.571 & 0.557 & 0.544 & 0.526 & 0.503 \\
           & $\pm0.012$ &  $\pm0.013$ & $\pm0.014$ & $\pm0.014$ & $\pm0.016$ \\ 
\hline
2.0 -- 2.5 & 0.551 & 0.540 & 0.527 & 0.507 & 0.487 \\
           & $\pm0.013$ & $\pm0.014$ &  $\pm0.014$ & $\pm0.016$ & $\pm0.018$ \\
\hline
\end{tabular}
\end{center}
}  
\caption{Multiplicity correlations for events at $\sqrt{s} = 7$~TeV
for events with a minimum of two charged particles in the kinematic
interval $p_{\rm T} > 100$~MeV and $|\eta| < 2.5$ for different
combinations of forward and backward pseudorapidity interval.  The
quoted uncertainties are the systematic uncertainties; the statistical
uncertainties are approximately $\pm0.001$ or less and are small by
comparison.  }
\label{tab:matrix}
\end{sidewaystable}

\begin{sidewaystable}[!ht]
{
\begin{center}
\begin{tabular}{|c|c|c|c|c|c||c|}
\hline
$|\eta|$ interval& $0.0-0.5$ & $0.5-1.0$  & $1.0-1.5$  & $1.5-2.0$  & $ 2.0-2.5$ & $0.0-2.5$  \\
\hline
900~GeV  & 0.498 & 0.407 & 0.359 & 0.310  & 0.259  & 0.652 \\
\vstrut & \B${\pm0.001}\,\,{\pm0.011}$\B & \B${\pm0.001}\,\,{\pm0.011}$ \B & \B${\pm0.002}\,\,{\pm0.013}$ \B & \B${\pm0.002}\,\,{\pm0.015}$ \B &\B ${\pm0.002}\,\,{\pm0.015}$ \B & \B${\pm0.001}\,\,{\pm0.007}$\B \\
\hline
7~TeV     & 0.666  & 0.596  & 0.560 & 0.526  & 0.487 & 0.801 \\
\vstrut & \B$\pm0.011$\B & \B $\pm0.011$ \B & \B $\pm$0.013  \B & \B $\pm0.014$   \B & \B $\pm0.018$  \B & \B $\pm0.008$ \B\\
\hline
\end{tabular}
\end{center}
} 
\caption{Forward-backward multiplicity correlation for 
charged particles in symmetrically opposite
$\eta$-intervals for events with a minimum of two charged particles in
the kinematic interval $p_{\rm T} > 100$~MeV and $|\eta| < 2.5$ at
$\sqrt{s}=$900~GeV and 7~TeV. The final column shows the global
result.  The quoted uncertainties are first statistical, second
systematic;
 where the statistical uncertainty is less than approximately 
$\pm0.001$, only the systematic uncertainty is shown.}
\label{tab:mult} 
\end{sidewaystable}

\begin{sidewaystable}[!ht]
{
\begin{center}
\begin{tabular}{|c|c|c|c|c|c||c|}
\hline
\hspace*{1.1ex} $|\eta|$ interval\B\B& $0.0-0.5$ & $0.5-1.0$ & $1.0-1.5$ & $1.5-2.0$ & $2.0-2.5$ & $0.0-2.5$ \\ 
\!\!$p_{\rm {T min}}$ [GeV] \!& & & & & & \\ 
\hline
0.1 & 0.666 & 0.596  & 0.560 & 0.526  & 0.487 & 0.801 \\
\vstrut & \B $\pm0.011$\B & \B $\pm0.011$\B & \B $\pm$ 0.013 \B & \B $\pm 0.014$  \B & \B $\pm.018$\B & \B $\pm0.008$\B  \\
\hline
0.2  & 0.639 	& 0.565	& 0.534 	& 0.498	& 0.458 & 0.790 \\
\vstrut& $\pm0.009$\B & \B  $\pm0.010$\B & \B  $\pm0.011$\B & \B $\pm0.012$\B & \B $\pm0.017$\B & \B $\pm0.008$\B \\
\hline
0.3  & 0.602 & 0.529	& 0.500	& 0.462	& 0.426  & 0.775 \\
\vstrut \B & \B $\pm0.009$ \B & $\pm0.009$\B & \B $\pm0.011$ \B & $\pm0.012$\B & \B $\pm0.016$ \B & \B $\pm0.008$\B \\
\hline
0.5 \B & \B  0.518 & 0.445	& 0.416	& 0.377 & 0.341 & 0.726 \\
\vstrut	& $\pm0.009$\B & \B  $\pm0.009$ \B & \B  $\pm0.010$ \B & \B $\pm0.014$ \B & \B $\pm0.017$ \B & \B $\pm0.010$\B   \\
\hline
1.0 & 0.291	& 0.216 	& 0.190	& 0.158	& 0.132 &  0.506 \\
\vstrut& $\pm0.010$\B & \B  $\pm0.011$ \B & \B $\pm0.013$ \B & \B $\pm0.018$\B & \B $\pm0.017$\B & \B $\pm0.023$\B  \\
\hline
1.5 &	0.128 	& 0.063 	& 0.047	& 0.026	& 0.010  & 0.234 \\
\vstrut & \B ${\pm0.001}\,\,{\pm0.006}$\B & \B ${\pm0.001}\,\,{\pm0.008}$\B & \B ${\pm0.001}\,\,{\pm0.016}$\B & \B ${\pm0.001}\,\,{\pm0.012}$ \B & 
\B ${\pm0.001}\,\,{\pm0.014}$\B & \B $\pm0.033$\B  \\
\hline
2.0 &	 0.036 	& --0.020 	& --0.031	& --0.041	& --0.046  & --0.008 \\
\vstrut & \B ${\pm0.001}\,\,{\pm0.006}$\B & \B ${\pm0.001}\,\,{\pm0.008}$\B & \B ${\pm0.001}\,\,{\pm0.007}$\B & \B ${\pm0.001}\,\,{\pm0.005}$ \B & \B ${\pm0.001}\,\,{\pm0.007}$ \B & \B${\pm0.001}\,\,{\pm0.019}$\B \\
\hline
\end{tabular}
\end{center}
} 
\caption{Forward-backward charged-particle multiplicity correlations
at $\sqrt{s} = 7$~TeV, in symmetrically opposite $\eta$-intervals, for
events with at least two charged particles with $|\eta| < 2.5$ and
$p_T > p_{\rm {T min}}$, for varying values of $p_{\rm {T min}}$. The
final column shows the global result.  The quoted uncertainties are
first statistical, second systematic; 
 where the statistical uncertainty is less than approximately
$\pm0.001$, only the systematic uncertainty is shown.}
\label{tab:manypt}
\end{sidewaystable}

\begin{sidewaystable}[!ht]
{
\begin{center}
\begin{tabular}{|l|c|c|c|c|c||c|   }
\hline
\vstrut$|\eta|$ interval \B & $0.0-0.5$ & $0.5-1$  & $1.0-1.5$  & $1.5-2.0$  & $ 2.0-2.5$ & $0.0-2.5$  \\
\hline
900~GeV  & 0.445 & 0.362 & 0.328 & 0.284  & 0.236 &0.637 \\
\vstrut & \B ${\pm0.001}\,\,{\pm0.008}$ \B&\B ${\pm0.002}\,\,{ \pm0.009}$ \B&\B ${\pm0.002}\,\,{ \pm0.010}$ \B&\B ${\pm0.002}\,\,{ \pm0.013}$ \B
&\B ${\pm0.003}\,\,{ \pm0.009}$ \B&\B${\pm0.001}\,\,{\pm0.009}$\,\B \\
\hline
7~TeV     & 0.590  & 0.519  & 0.492 & 0.455  & 0.412 & 0.770\\
\vstrut & $\pm0.010$ & $\pm0.009$ & $\pm$0.010  & $\pm0.011$ & $\pm0.016$ & 
 $\,^{+0.007}_{-0.008}$\\
\hline
\end{tabular}
\end{center}
} 
\caption{Forward-backward summed \pT\ correlation for charged
particles in symmetrically opposite $\eta$ intervals, for events with
at least two charged particles in the kinematic interval $p_{\rm T} >
100$~MeV and $|\eta| < 2.5$ at $\sqrt{s}=$900~GeV and 7~TeV, for
corrected data.  The quoted uncertainties are first statistical,
second systematic; where the statistical uncertainty less than approximately 
$\pm0.001$, only the systematic uncertainty is shown.  }
\label{tab:sumpt} 
\end{sidewaystable}

\newpage 

\clearpage
\section{Conclusions}
\label{sec:conclusions}

The forward-backward correlation in charged-particle multiplicity
and in summed charged-particle \pT\ has been measured at ATLAS in
minimum-bias events at $\sqrt{s}=900$~GeV and 7~TeV.  This is the
first measurement of summed-\pT\ correlations. Both types of
correlation fall in strength with increasing separation in pseudorapidity.
The multiplicity correlation is found to differ by up to
about 15\% from some of the standard Monte Carlo tunes at 7~TeV, and
by somewhat more at 900~GeV; the shapes of the distributions 
are generally similar but some of the tunes show systematically different
trends from the data. With the summed-\pT\ correlation, similar
conclusions are found.  It should be noted that present Monte Carlo
generators have not been tuned using these correlations.

Both types of correlation are found to be significantly stronger at
7~TeV than at 900~GeV, the relative increase being greater at 
larger intervals in pseudorapidity. This trend is reproduced in most of the MC
models. As expected, a  strong tendency is seen for the multiplicity
correlations to fall with increasing minimum transverse momentum of the
charged particles.  This illustrates well the transition between the
soft, non-perturbative regime of parton string or cluster fragmentation and the
jet-dominated regime of perturbative quantum chromodynamics.

Azimuthal charged-particle distributions relative to the leading
charged particle have been measured by means of two variables, using
the full pseudorapidity reach of the ATLAS tracking detectors.  These variables minimise
inter-jet effects in different ways, and give different sensitivities
to the jet-like properties of the event distributions.  In both cases,
the MC models give fair to good descriptions of the data, provided
that the selected pseudorapidity range is within one unit of
zero. As the range in pseudorapidity is increased, however, disagreements
with most of the models become more pronounced, in describing both the
jet-influenced region near the leading charged particle and also the
opposite-side region. This indicates the complexity of the effects that the
models are seeking to describe when broad features of the  events are 
being considered.

The event features studied here provide new means of
discriminating between the different possible tunes of the Monte Carlo
models. They will need to be taken into account, if the most accurate
description of particle processes at the highest LHC energies is to be
attained.
 
\section*{Acknowledgements}
We thank CERN for the very successful operation of the LHC, as well as the
support staff from our institutions without whom ATLAS could not be
operated efficiently.

We acknowledge the support of ANPCyT, Argentina; YerPhI, Armenia; ARC,
Australia; BMWF, Austria; ANAS, Azerbaijan; SSTC, Belarus; CNPq and FAPESP,
Brazil; NSERC, NRC and CFI, Canada; CERN; CONICYT, Chile; CAS, MOST and NSFC,
China; COLCIENCIAS, Colombia; MSMT CR, MPO CR and VSC CR, Czech Republic;
DNRF, DNSRC and Lundbeck Foundation, Denmark; EPLANET and ERC, European Union;
IN2P3-CNRS, CEA-DSM/IRFU, France; GNAS, Georgia; BMBF, DFG, HGF, MPG and AvH
Foundation, Germany; GSRT, Greece; ISF, MINERVA, GIF, DIP and Benoziyo Center,
Israel; INFN, Italy; MEXT and JSPS, Japan; CNRST, Morocco; FOM and NWO,
Netherlands; RCN, Norway; MNiSW, Poland; GRICES and FCT, Portugal; MERYS
(MECTS), Romania; MES of Russia and ROSATOM, Russian Federation; JINR; MSTD,
Serbia; MSSR, Slovakia; ARRS and MVZT, Slovenia; DST/NRF, South Africa;
MICINN, Spain; SRC and Wallenberg Foundation, Sweden; SER, SNSF and Cantons of
Bern and Geneva, Switzerland; NSC, Taiwan; TAEK, Turkey; STFC, the Royal
Society and Leverhulme Trust, United Kingdom; DOE and NSF, United States of
America.

The crucial computing support from all WLCG partners is acknowledged
gratefully, in particular from CERN and the ATLAS Tier-1 facilities at
TRIUMF (Canada), NDGF (Denmark, Norway, Sweden), CC-IN2P3 (France),
KIT/GridKA (Germany), INFN-CNAF (Italy), NL-T1 (Netherlands), PIC (Spain),
ASGC (Taiwan), RAL (UK) and BNL (USA) and in the Tier-2 facilities
worldwide.


\bibliography{references}
\appendix
\section{Summary of Monte Carlo tunes}
\label{app:a}

Table~\ref{tab:MCTunes} summarises the main features of the \PYTHIA- and
\HERWIG-based Monte Carlo tunes employed in the analysis and referred
to in the text.\\[5mm]

\begin{table}[h]
\centering
\begin{tabular}{|c|c|c|c|c|}
\hline \hline
Generator	& Tune		& Shower & PDF& Additional information\\
	& 		& ordering &  & \\
\hline \hline
\multirow{4}{*}{\PYTHIA~6.4}& DW& {\!virtuality\!}& CTEQ 5L& Non-interleaved MPI model\\
\cline{2-5}
& MC09& $\pT$& MRSTLO*& MPI interleaved with ISR\\
\cline{2-5}
& Perugia 2011& $\pT$& CTEQ 5L& MPI interleaved with ISR\\
\cline{2-5}
& AMBT2B& $\pT$& CTEQ 6L1& MPI interleaved with ISR\\
\hline
\multirow{3}{*}{\PYTHIA~8.150}& \multirow{3}{*}{4C}& \multirow{3}{*}{$\pT$}& \multirow{3}{*}{CTEQ 6L1} & MPI interleaved with ISR\\
&&&& and FSR; updated \\
&&&& model for diffraction \\
\hline
\multirow{4}{*}{{\sc herwig}{\small++} 2.5}& \multirow{2}{*}{MU900-2}& \multirow{2}{*}{angle}& MRST& MB tune for 900~GeV only; \\
&&&2007LO*& no model for diffraction\\
\cline{2-5}
& \multirow{2}{*}{UE7-2}& \multirow{2}{*}{angle}& MRST & UE tune for 7~TeV only;\\
&&&2007LO*& no model for diffraction\\
\hline \hline
\end{tabular}
\caption{Main features of the different \PYTHIA\ and \HERWIG\ tunes
used in the comparisons with the data, as used with
CTEQ~\cite{ref:cteq5,ref:cteq6} and MRST~\cite{ref:mrst} parton
distribution functions.}
\label{tab:MCTunes}
\centering
\end{table}
\clearpage
\begin{flushleft}
{\Large The ATLAS Collaboration}

\bigskip

G.~Aad$^{\rm 48}$,
B.~Abbott$^{\rm 110}$,
J.~Abdallah$^{\rm 11}$,
A.A.~Abdelalim$^{\rm 49}$,
A.~Abdesselam$^{\rm 117}$,
O.~Abdinov$^{\rm 10}$,
B.~Abi$^{\rm 111}$,
M.~Abolins$^{\rm 87}$,
O.S.~AbouZeid$^{\rm 157}$,
H.~Abramowicz$^{\rm 152}$
% ...
\bigskip

% ...
$^{10}$ Institute of Physics, Azerbaijan Academy of Sciences, Baku, Azerbaijan\\
$^{11}$ Institut de F\'isica d'Altes Energies and Departament de F\'isica de la Universitat Aut\`onoma  de Barcelona and ICREA, Barcelona, Spain\\
% ...
$^{48}$ Fakult\"{a}t f\"{u}r Mathematik und Physik, Albert-Ludwigs-Universit\"{a}t, Freiburg i.Br., Germany\\
$^{49}$ Section de Physique, Universit\'e de Gen\`eve, Geneva, Switzerland\\
% ...
$^{87}$ Department of Physics and Astronomy, Michigan State University, East Lansing MI, United States of America\\
% ...
$^{110}$ Homer L. Dodge Department of Physics and Astronomy, University of Oklahoma, Norman OK, United States of America\\
$^{111}$ Department of Physics, Oklahoma State University, Stillwater OK, United States of America\\
% ...
$^{117}$ Department of Physics, Oxford University, Oxford, United Kingdom\\
% ...
$^{152}$ Raymond and Beverly Sackler School of Physics and Astronomy, Tel Aviv University, Tel Aviv, Israel\\
% ...
$^{157}$ Department of Physics, University of Toronto, Toronto ON, Canada\\
% ...
\end{flushleft}

\end{document}